\documentclass[preprint,aps,amsmath,amssymb,nofootinbib,12pt]{revtex4}

\usepackage{epsfig}
\usepackage{slashed}
\usepackage{graphicx}
\usepackage{multirow,color}
\usepackage{amsmath}
\usepackage{float}
\usepackage{diagbox}
\usepackage{CJK}
\usepackage{color}
\usepackage{xcolor}
\usepackage{times}
\usepackage{subfigure}
\usepackage{bm}
\usepackage{braket}
\usepackage{booktabs}
\usepackage{array}
\usepackage[mathscr]{euscript}
\usepackage{caption}
\usepackage{makecell}
\usepackage{epstopdf}
\usepackage{hyperref}

\makeatletter

\newcommand{\Rmnum}[1]{\expandafter\@slowromancap\romannumeral #1@}
\makeatother
\graphicspath{{fig/}}
\begin{document}
\title{Prospects for probing light photophobic axion-like particles via displaced vertex signals at the CEPC}
\author{Chong-Xing Yue$^{1,2}$}
\thanks{cxyue@lnnu.edu.cn}
\author{Xin-Yang Li$^{1,2}$}
\thanks{lxy91108@163.com~(Corresponding author)}

\affiliation{
$^1$Department of Physics, Liaoning Normal University, Dalian 116029, China\\
$^2$Center for Theoretical and Experimental High Energy Physics, Liaoning Normal University,  Dalian 116029, China}

\begin{abstract}
In recent years, long-lived particles~(LLPs) have attracted increasing attention in searches for physics beyond the Standard Model~(SM). In this paper, we investigate the discovery prospects for light, long-lived ALPs predicted by the photophobic ALP scenario through displaced-vertex signals at the CEPC, with a center-of-mass energy of $\sqrt{s}=91.2 $ GeV and an integrated luminosity of $\mathcal{L}=$ $100$ ab$^{-1}$. After comparing several possible single-production processes for the photophobic ALP, we focus on the dominant process $e^+ e^- \to Z \to a \gamma $, in which the ALP $a$ subsequently decays into a pair of charged leptons at a displaced vertex. Dedicated Monte Carlo simulations are performed for the $\mu^+ \mu^- \gamma$ and $\tau_h^+ \tau_h^- \slashed{E}_T \gamma$ signals. For the $\mu^+\mu^-\gamma$ signal, the CEPC is sensitive to the parameter region $g_{aWW} \in [1.27\times10^{-3},6.80\times10^{-1}]~\mathrm{TeV}^{-1}$ for $m_a \in [1,4]~\mathrm{GeV}$. For the $\tau_h^+\tau_h^-\slashed{E}_T\gamma$ signal, the accessible region is $g_{aWW} \in [7.00\times10^{-4},9.40\times10^{-3}]~\mathrm{TeV}^{-1}$ for $m_a \in [4,9]~\mathrm{GeV}$. These results demonstrate the substantial potential of the CEPC to explore light, long-lived ALPs through displaced-vertex signals, providing complementary coverage to existing searches at LEP and the LHC, as well as to the projected reach of the HL-LHC.
\end{abstract}
\maketitle

\section{Introduction}

The origin of neutrino masses~\cite{Gonzalez-Garcia:2007dlo}, the nature of dark matter~\cite{Bertone:2004pz,Dienes:2011ja}, the strong CP problem~\cite{Peccei:1977hh,Peccei:1977ur,Wilczek:1977pj}, and the hierarchy problem~\cite{Feng:2013pwa} remain among the deepest mysteries in particle physics. Over the past decades, many experimental efforts addressing these puzzles have focused on new particles with sizeable couplings to the Standard Model~(SM) and masses above several hundred GeV or even at the TeV scale. More recently, lighter feebly interacting particles~(FIPs)~\cite{Lanfranchi:2020crw,Antel:2023hkf} with masses at the GeV scale have emerged as a well-motivated and complementary possibility. Owing to their extremely weak interactions with SM particles, these FIPs are typically long-lived and may decay at macroscopic distances from their production points. When such decays occur within a detector, their visible decay products can be reconstructed as displaced vertices~(DVs). Consequently, conventional searches targeting prompt signals often lose sensitivity, whereas displaced vertex signals provide a promising avenue for detecting long-lived particles~(LLPs)~\cite{Jeanty:2025wai}.

Searches for LLPs have been extensively pursued at the LHC. The ATLAS, CMS, and LHCb Collaborations have conducted numerous analyses based on displaced vertex signals~(see, e.g., Refs.~\cite{ATLAS:2018rjc,ATLAS:2019fwx,ATLAS:2023oti,ATLAS:2020xyo,ATLAS:2019tkk,ATLAS:2018tup,ATLAS:2019jcm,
ATLAS:2022gbw,ATLAS:2022zhj,ATLAS:2021jig,ATLAS:2018pvw,CMS:2022qej,CMS:2021sch,CMS:2021tkn,CMS:2020iwv,CMS:2021yhb,LHCb:2017xxn,LHCb:2016buh}). While the LHC demonstrates strong sensitivity across a broad range of LLP scenarios, future lepton colliders such as the CEPC~\cite{CEPCStudyGroup:2018ghi,CEPCStudyGroup:2023quu} and FCC-ee~\cite{FCC:2018evy,FCC:2025lpp} can provide complementary and promising platforms for LLP searches. Operating as $Z$-boson and Higgs-boson factories, these facilities are expected to produce up to $4.1\times10^{12}$ $Z$ bosons and $4.3\times10^{6}$ Higgs bosons~\cite{Gao:2022lew}, enabling unprecedented electroweak and Higgs precision measurements. Moreover, the clean experimental environment and high integrated luminosity of these colliders are expected to reduce backgrounds and enhance sensitivity to LLPs. Motivated by these advantages, numerous studies~\cite{Cheung:2019qdr,Wang:2019orr,Alipour-Fard:2018lsf,Cao:2023smj,Suarez:2021hpn,Zhang:2024bld,
Blondel:2022qqo,Urquia-Calderon:2023dkf,Chrzaszcz:2021nuk} have explored the phenomenology of LLPs at future lepton colliders, including vector-like leptons~(VLLs) and axion-like particles~(ALPs).

ALPs~\cite{Kim:1979if,Shifman:1979if,Dine:1981rt,Zhitnitsky:1980tq,Anselm:1981aw,Ringwald:2014vqa} are among the most well-motivated candidates for feebly interacting particles. They are pseudo-Nambu-Goldstone bosons associated with the spontaneous breaking of global $U(1)$ symmetries and appear as singlet pseudoscalars under the SM gauge groups. Over the years, a wide range of new physics scenarios have predicted the existence of ALPs. Of particular interest is the photophobic ALP scenario~\cite{Craig:2018kne}, in which the ALP has no tree-level couplings to SM photons or fermions and interacts only with other electroweak gauge bosons~($WW,ZZ,Z\gamma$) at tree level. Although direct couplings to photons and fermions are absent, ALP interactions with electroweak gauge bosons can induce effective ALP-fermion and ALP-photon couplings at the loop level. Most existing phenomenological studies have focused on heavy photophobic ALPs and their prompt-decay signals~\cite{Aiko:2024xiv,Ding:2024djo,Mao:2024kgx,Feng:2025kof,Ding:2025wvb}. However, the ALPs predicted by this scenario can be long-lived, particularly in the low-mass region. When the photophobic ALP mass lies below the threshold for on-shell decays into massive electroweak gauge bosons, these channels are kinematically forbidden. Light photophobic ALPs then decay predominantly through loop-induced channels into fermions and photons, leading to a suppressed total width and an enhanced proper decay length. As a result, light photophobic ALPs can manifest as LLPs, giving rise to distinctive signals at high-energy colliders. Motivated by this feature, several studies have explored the phenomenology of long-lived photophobic ALPs. For instance, Ref.~\cite{Wang:2025ncc} investigated prompt-lepton and mono-photon signals of these ALPs at future lepton colliders. However, when the ALP decay length becomes comparable to the detector scale, displaced vertex searches provide a powerful and complementary probe, accessing regions of parameter space that are not efficiently covered by prompt-decay and missing-energy searches. Future lepton colliders, such as the CEPC, offer a promising opportunity to probe light long-lived photophobic ALPs via displaced vertex signals, owing to their clean experimental environments, high integrated luminosities, and precise measurement capabilities.

In this paper, we investigate the prospects for probing light long-lived photophobic ALPs via displaced
vertex signals within the main tracking system of the CEPC at the center-of-mass energy $\sqrt{s}=91.2 $ GeV and with an integrated luminosity of $\mathcal{L}=$ $100$ ab$^{-1}$. We consider the production of a photophobic ALP in association with a photon $\gamma$, a lepton pair, or a jet pair. Among these production processes, photon-associated production has the largest cross section. We therefore focus on the process $e^+ e^- \to Z \to a \gamma $ as a representative example, where the ALP $a$ subsequently decays into a pair of displaced charged leptons. Our analysis demonstrates that the CEPC can probe regions of parameter space that remain unexplored by current and future collider experiments, providing sensitivity complementary to previous searches.

The paper is organized as follows. In Sec. II, we introduce the theoretical framework of photophobic ALPs, discuss their possible decay modes and lifetimes, and analyze their single production at the CEPC. In Sec. III, we present a detailed analysis of the possibility of detecting light long-lived photophobic ALPs via the process $e^+ e^- \to Z \to a \gamma $, with the ALP $a$ subsequently decaying into a pair of displaced charged leptons within the main tracking system of the CEPC. In Sec. IV, we summarize our conclusions and discussion.

\section{The theory framework}

We consider the photophobic ALP scenario~\cite{Craig:2018kne}, in which the ALP interacts only with the $SU(2)_L$ and $U(1)_Y$ gauge bosons, while tree-level ALP couplings to fermions and gluons are absent. This setup is consistent with the ultraviolet~(UV) boundary conditions at the high scale $\Lambda$. The effective Lagrangian before electroweak symmetry breaking can be written as~\cite{Georgi:1986df}

\begin{equation}
\begin{aligned}
\mathcal{L}_{\rm ALP} &=
\frac{1}{2}\partial_{\mu}a\partial^{\mu}a
-\frac{1}{2} m_{a}^{2} a^{2} -
c_{\tilde{W}} \frac{a}{f_a} W^{i}_{\mu\nu} \tilde{W}^{\mu\nu,i} - c_{\tilde{B}} \frac{a}{f_a} B_{\mu\nu} \tilde{B}^{\mu\nu},
\label{eq:1}
\end{aligned}
\end{equation}
where $W^{i}_{\mu\nu}$ and $B_{\mu\nu}$ represent the gauge field strength tensors associated with the $SU(2)_L$ and $U(1)_Y$ gauge groups, respectively. The corresponding dual field strength tensors are defined as $\tilde{V}^{\mu\nu}=\frac{1}{2}\epsilon^{\mu\nu\lambda\kappa}V_{\lambda\kappa}$ $(V = W^{i}$, $B)$, with
$\epsilon^{0123} = 1$. The ALP mass and decay constant are denoted by $m_a$ and $f_a$, respectively. The coefficients $c_{\tilde{W}}/{f_a}$ and $c_{\tilde{B}}/{f_a}$, together with $m_a$, are treated as independent free parameters.

After electroweak symmetry breaking, the interactions of the ALP with the electroweak gauge bosons can be written as
\begin{equation}
\begin{aligned}
\mathcal{L}_{\mathrm{ALP}}
&= -\frac{1}{4}g_{a\gamma\gamma}aF_{\mu\nu}\tilde{F}^{\mu\nu}
-\frac{1}{2}g_{a\gamma Z}aZ_{\mu\nu}\tilde{F}^{\mu\nu}
-\frac{1}{4}g_{aZZ}aZ_{\mu\nu}\tilde{Z}^{\mu\nu}
- \frac{1}{2} g_{aWW} \,a\,W_{\mu \nu}^+ \tilde{W}^{\mu \nu,-},
\label{eq:2}
\end{aligned}
\end{equation}
where the corresponding coupling constants are given by
\begin{equation} \label{eq:3}
    \begin{aligned}
       g_{a\gamma\gamma} &= \frac{4}{f_{a}}(s_{W}^{2}c_{\tilde{W}}+c_{W}^{2}c_{\tilde{B}}), ~~~~~~~
g_{aZ\gamma} = \frac{4}{f_{a}}s_{W}c_{W}(c_{\tilde{W}}-c_{\tilde{B}}), \\
g_{aZZ} &= \frac{4}{f_{a}}(c_{W}^{2}c_{\tilde{W}}+s_{W}^{2}c_{\tilde{B}}), ~~~~~~~
g_{aWW} = \frac{4}{f_{a}}c_{\tilde{W}},
    \end{aligned}
\end{equation}
with $s_W$ and $c_W$ denoting the sine and cosine of the Weinberg angle, respectively. In the photophobic ALP scenario, the tree-level coupling of the ALP to photons vanishes as a consequence of the UV boundary conditions~(see Ref.~\cite{Craig:2018kne} for details). Accordingly, $g_{a\gamma\gamma}=\frac{4}{f_{a}} (s_{W}^{2} c_{\tilde{W}}+c_{W}^{2} c_{\tilde{B}}) = 0$, which implies the relation $s_{W}^{2} c_{\tilde{W}}+c_{W}^{2} c_{\tilde{B}} = 0$. Consequently, the couplings $g_{aZ\gamma}$, $g_{aZZ}$, and $g_{aWW}$ are not independent and can be expressed as

\begin{equation}\label{eq:4}
g_{aZ\gamma} = \frac{s_{W}}{c_{W}}g_{aWW},\quad
g_{aZZ} = \frac{c_{2W}}{c_{W}^{2}}g_{aWW}.
\end{equation}
Therefore, the photophobic ALP scenario is characterized by the ALP mass $m_a$ and the coupling $g_{aWW}$.

Owing to the UV boundary conditions, tree-level ALP--fermion interactions vanish. However, an effective ALP--fermion coupling can be generated through one-loop renormalization group evolution~(RGE) from the UV scale $\Lambda$ down to the electroweak scale and can be written as~\cite{Bauer:2017ris,Bauer:2020jbp}
\begin{equation}\label{eq:5}
\begin{aligned}
    g_{aff}^{\rm eff} &=
    \frac{3}{4s_{W}^{2}}\frac{\alpha}{4\pi}g_{aWW} \ln{\frac{\Lambda^{2}}{m_{W}^{2}}}
     \\ &\quad
    +\frac{3}{c_{W}^2}\frac{\alpha}{4\pi}g_{aWW}Q_{f}
        (I_{3}^{f}-2Q_{f}s_{W}^{2})
        \ln{\frac{\Lambda^{2}}{m_{Z}^{2}}}
     \\ &\quad
    +\frac{3 c_{2W}}{s_{W}^{2}c_{W}^{4}}\frac{\alpha}{4\pi}g_{aWW}
        (Q_{f}^{2}s_{W}^{4}-I_{3}^{f}Q_{f}s_{W}^{2}+\frac{1}{8})
        \ln{\frac{\Lambda^{2}}{m_{Z}^{2}}}.
\end{aligned}
\end{equation}
Here, $Q_{f}$ and $I_{3}^{f}$ denote the electric charge and the third component of weak isospin of the fermion $f$, respectively, and $\alpha = e^2/(4\pi)$ is the fine-structure constant. In this paper, the cutoff scale is taken to be $4 \pi f_a= 4 \pi \times 10$ TeV. In the above expression, we retain only the terms containing $\ln(\Lambda^2/m_V^2)$ (with $V = Z, W$), which are logarithmically enhanced. The subleading terms are given in Refs.~\cite{Bauer:2017ris,Bauer:2020jbp,Bonilla:2021ufe}; however, they are numerically small and can be safely neglected in our analysis.

The couplings of the ALP to the electroweak gauge bosons can induce a small effective ALP-photon coupling at one loop~\cite{Bauer:2017ris}, which can be expressed as
\begin{equation}\label{eq:5}
    g_{a\gamma\gamma}^{\rm eff} = \frac{2\alpha}{\pi} g_{aWW}B_{2}(\tau_{W}),
\end{equation}
where the loop function $B_{2}$ is defined as
\begin{equation}\label{eq:6}
    B_{2}(\tau) = 1-(\tau-1)f^{2}(\tau),\quad \text{with}~f(\tau) =
    \begin{cases}
        \arcsin{\frac{1}{\sqrt{\tau}}}, & \text{for } \tau \geq 1 \\
        \frac{\pi}{2} + \frac{i}{2} \log{\frac{1+\sqrt{1-\tau}}{1-\sqrt{1-\tau}}}, & \text{for } \tau < 1
    \end{cases},
\end{equation}
with $\tau_{W}=4m_{W}^{2}/m_{a}^{2}$.

Under the UV boundary conditions specified above, the ALP does not couple directly to gluons. However, the effective ALP-fermion coupling $g_{aff}^{\rm eff}$ can induce an effective ALP-gluon coupling at one loop. Because $g_{aff}^{\rm eff}$ itself arises from ALP interactions with electroweak gauge bosons at one loop, the resulting ALP-gluon coupling is generated at the two-loop level and can be written as~\cite{Aiko:2024xiv}
\begin{equation}\label{eq:8}
g_{a G G}^{\mathrm{eff}} = \frac{1}{32 \pi^{2}} \sum_{f = u, d, s} g_{a f f}^{\mathrm{eff}}.
\end{equation}
Since this coupling arises only at the two-loop level and is therefore highly suppressed, its contribution to the signal processes considered in this study is negligible and can be safely ignored.

As discussed above, the ALP can exhibit various decay modes, depending on its mass and the corresponding kinematic thresholds. The expressions for the corresponding partial decay widths are given by~\cite{Bauer:2017ris,Bauer:2020jbp,Bonilla:2021ufe}

\begin{equation}\label{eq:9}
\begin{aligned}
 &
\Gamma(a\to\gamma\gamma) =  \frac{m_a^3}{64\pi}|g_{a\gamma\gamma}^\text{eff}|^2,
 \\
 &
  \Gamma(a\to f \bar{f}) =  N_{c}^f \frac{m_a m_{f}^2|g_{aff}^\text{eff}|^2}{8\pi}\sqrt{1-\frac{4m_{f}^2}{m_a^2}},
 \\
 &
  \Gamma(a\to Z \gamma) =  \frac{m_a^3 s_{W}^2}{32\pi c_{W}^2}|g_{aWW}|^2 (1-\frac{m_Z^2}{m_a^2})^3,
  \\
&
  \Gamma(a\to Z Z) =  \frac{m_a^3 c_{2W}^2}{64\pi c_{W}^4}|g_{aWW}|^2 (1-4\frac{m_Z^2}{m_a^2})^{3/2},
  \\
&
  \Gamma(a\to W^{+} W^{-}) =  \frac{m_a^3}{32\pi}|g_{a W W}|^2 (1-4\frac{m_W^2}{m_a^2})^{3/2}.
\end{aligned}
\end{equation}
Here, $c_{2W} \equiv \cos2\theta_W$, and $N_c^f = 1\ (3)$ denotes the color factor for leptons~(quarks).

When the ALP is lighter than the $Z$ boson, the three-body decay $a \to \gamma Z^{*} \to \gamma f\bar{f}$, mediated by an off-shell $Z$ boson, is also kinematically allowed. In the limit $m_a \ll m_Z$, the corresponding partial decay width is given by~\cite{Aiko:2024xiv}
\begin{equation}\label{eq:10}
\begin{aligned}
\Gamma(a \to \gamma Z^* \to \gamma f\bar{f}) = N_c^f \frac{g_Z^2 s_{W}^2 |g_{aWW}|^2}{30720\pi^3 c_{W}^2} \left[ (g_{V,f})^2 + (g_{A,f})^2 \right] \frac{m_a^7}{m_Z^4},
\end{aligned}
\end{equation}
where $g_Z = g/c_W$ is the $Z$-boson coupling constant, and $g_{V,f} = I_3^f - 2 Q_f s_W^2$ and $g_{A,f} = I_3^f$ denote the vector and axial-vector couplings of the $Z$ boson to fermions, respectively.
Owing to suppression from the off-shell $Z$-boson propagator and the limited three-body phase space, the decay width of the three-body channel remains several orders of magnitude smaller than that of the loop-induced two-body decay $a \to f\bar f$ throughout the parameter space considered in this work. Consequently, it has a negligible impact on the total ALP decay width and can be safely neglected in our phenomenological analysis.

\begin{figure}[H]
\begin{center}
\includegraphics [scale=0.4] {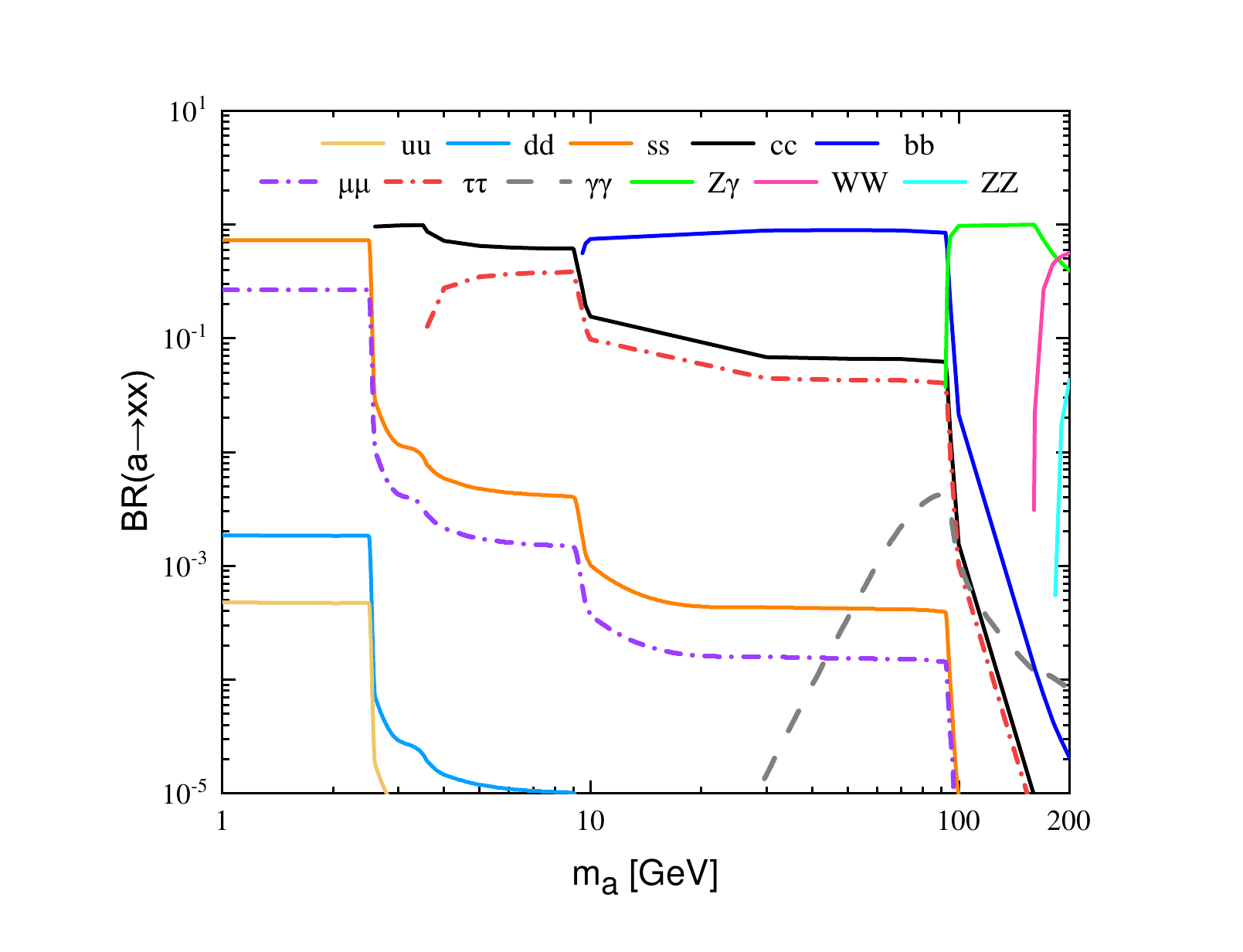}
\caption{Branching ratios for the different photophobic ALP decay modes as functions of the ALP mass $m_a$.}
\label{fig:1}
\end{center}
\end{figure}

The branching ratios for the dominant ALP decay modes are shown in Fig.~\ref{fig:1}. As shown in this figure, for relatively light ALPs, tree-level decays into massive electroweak gauge bosons are kinematically forbidden. In this mass range, light ALPs decay predominantly through loop-induced channels into fermions. More specifically, within the mass range $2m_\mu \leq m_a \leq 2m_c$, the dominant decay modes are $s \bar s$ and $\mu^+\mu^-$. For $2m_c \leq m_a \leq 2m_b$, the $c\bar{c}$ and $\tau^+\tau^-$ channels become dominant. For $m_a \geq 2m_b$, the $b\bar b$ channel dominates, while $c\bar c$ and $\tau^+\tau^-$ remain subdominant.
It should be noted that, owing to QCD confinement, ALP decays into light quarks cannot be directly observed as free quark final states; instead, they appear as hadronic final states. A quantitative description of these hadronic states requires nonperturbative methods, which are beyond the scope of this work (see Ref.~\cite{Bai:2025fvl} for details). Therefore, we adopt a parton-level treatment as a simplified approach to illustrate the ALP decay modes.

\begin{figure}[H]
\begin{center}
\includegraphics [scale=0.35] {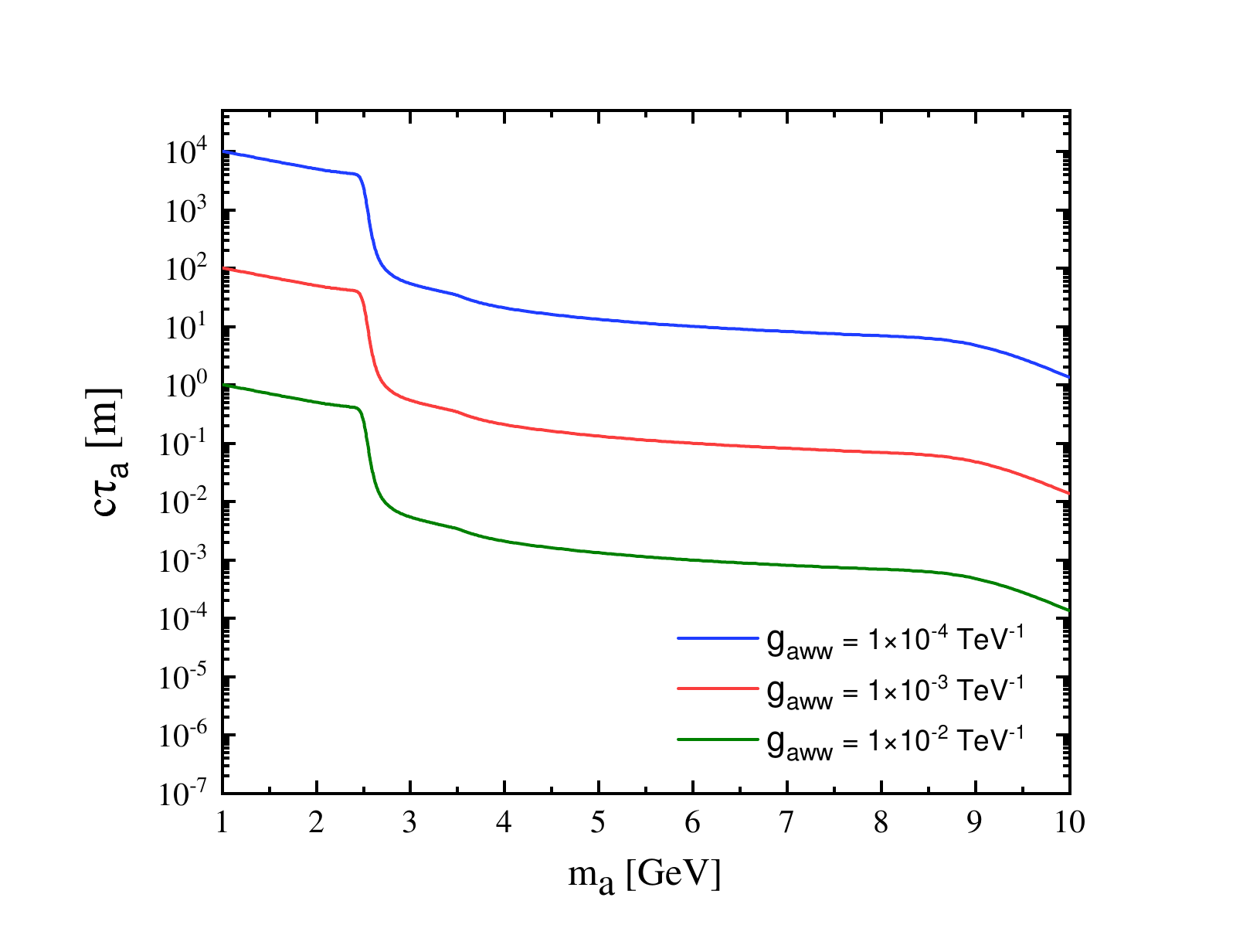}
\caption{The proper decay lengths of the ALP as functions of its mass, $m_a$, for several representative values of the coupling $g_{aWW}$.}
\label{fig:2}
\end{center}
\end{figure}

In this paper, we focus on ALPs with masses in the range of $1$--$10$ GeV and investigate their displaced-vertex signals in the photophobic ALP scenario. In this mass range, the ALP predominantly decays into fermions through loop-induced couplings. These couplings suppress the total decay width, allowing the ALP to behave as a long-lived particle and propagate a measurable distance inside the detector before decaying. The decay length in the laboratory frame is given by
\begin{equation}\label{eq:11}
L_a = \beta_a \gamma_a c \tau_a = \frac{c|\vec{p}_a|}{m_a \Gamma_a},
\end{equation}
where $\Gamma_a = 1/\tau_a$ is the total decay width of the ALP, and $\beta_a$ and $\gamma_a$ denote its velocity and Lorentz boost factor, respectively. The product $\beta_a\gamma_a = |\vec{p}_a|/m_a$, where $\vec{p}_a$ is the three-momentum of the ALP. The proper decay length $c\tau_a$ as a function of the ALP mass $m_a$ for several representative values of the coupling $g_{aWW}$ is shown in Figure~\ref{fig:2}. As illustrated in this figure, $c\tau_a$ depends strongly on both $m_a$ and $g_{aWW}$, increasing significantly as these parameters decrease. This wide range of proper decay lengths indicates that ALPs can be long-lived over a substantial region of the parameter space. Such long-lived ALPs can travel macroscopic distances before decaying within the detector, giving rise to distinctive DV signals that complement conventional ALP searches and can be efficiently explored within the main tracking system of the CEPC.

In general, displaced vertices can also be reconstructed in other detector components, such as the electromagnetic calorimeter~(ECAL), hadronic calorimeter~(HCAL), and muon spectrometer~(MS). In this work, we focus exclusively on the main tracking system for displaced-vertex reconstruction, because it provides precise vertex reconstruction for charged particles and is therefore well suited to identifying displaced decays occurring close to the interaction point. A dedicated study of displaced-vertex signals in the calorimeter systems would require specialized simulations and trigger strategies, which are beyond the scope of the present analysis. The MS is also not considered here, because its greater distance from the interaction point makes it more sensitive to particles with longer decay lengths and less sensitive to decays with relatively short lifetimes.

\section{The prospects for probing light photophobic long-lived ALPs at the CEPC}

\begin{figure}[H]
\begin{center}
\subfigure[]{\includegraphics [scale=0.65] {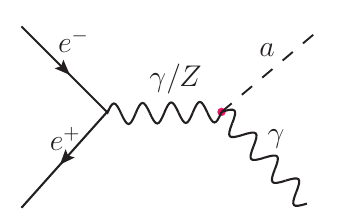}}
\subfigure[]{\includegraphics [scale=0.65] {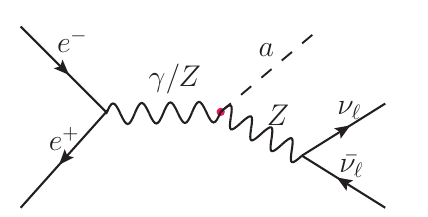}}
\subfigure[]{\includegraphics [scale=0.65] {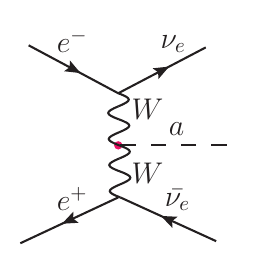}}\\
\subfigure[]{\includegraphics [scale=0.65] {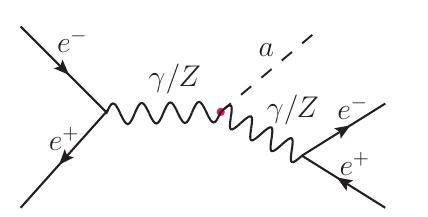}}
\subfigure[]{\includegraphics [scale=0.65] {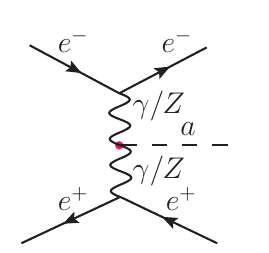}}
\subfigure[]{\includegraphics [scale=0.65] {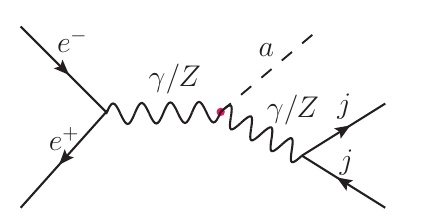}}
\caption{Feynman diagrams for the single-production processes of the photophobic ALP: $e^+ e^- \to a \gamma$~(a), $a \nu \bar{\nu}$~(b,c), $a e^+ e^-$~(d,e), and $a jj$~(f).}
\label{fig:3}
\end{center}
\end{figure}

\begin{figure}[H]
\begin{center}
\subfigure[]{\includegraphics [scale=0.3] {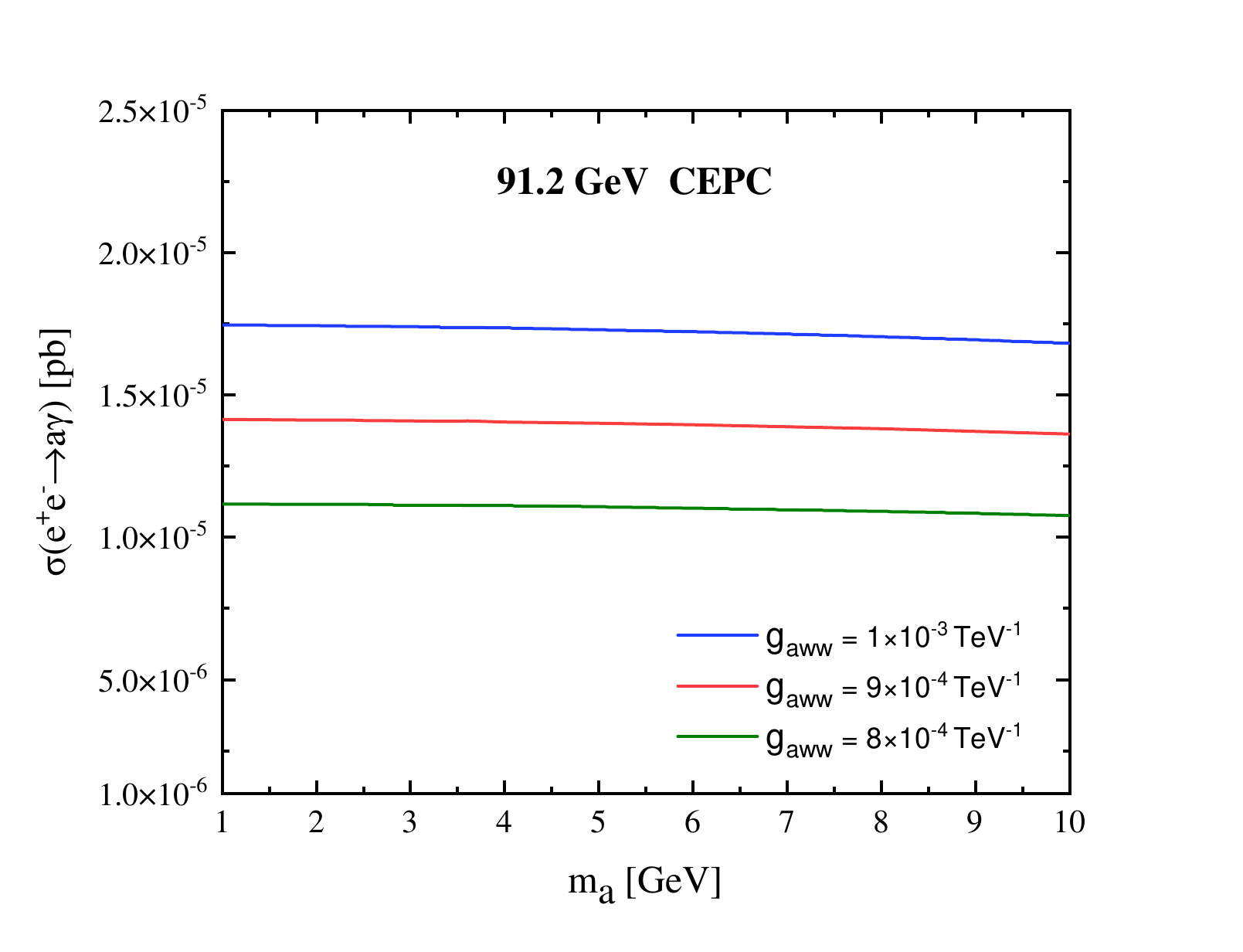}}
\hspace{-0.3in}
\subfigure[]{\includegraphics [scale=0.3] {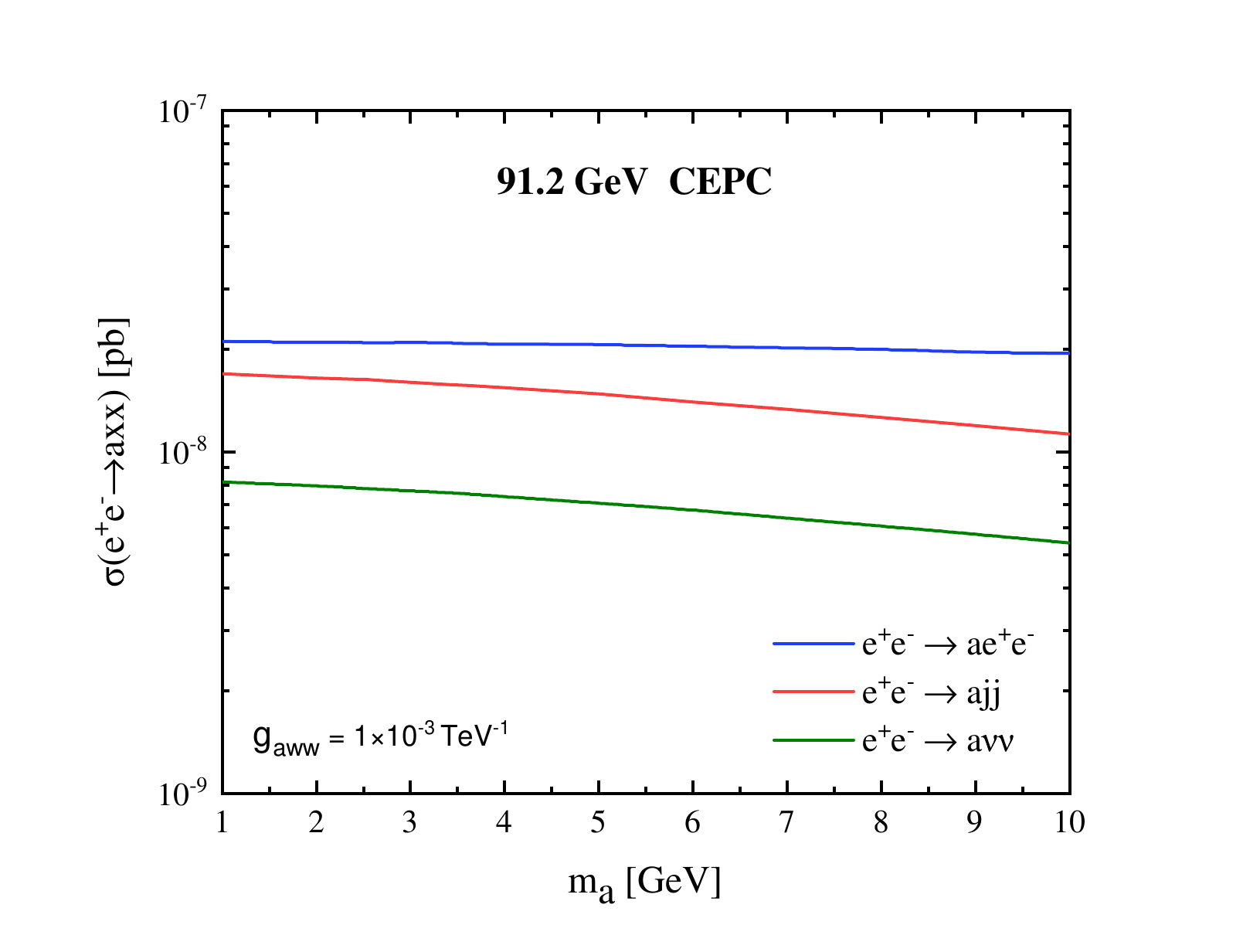}}
\caption{Production cross sections for the processes $e^+ e^- \to a \gamma$~(a), $a \nu \bar{\nu}$, $a e^+ e^-$, and $a jj$~(b) as functions of the ALP mass $m_a$ for different values of the coupling $g_{aWW}$ at the $91.2$ GeV CEPC.}
\label{fig:4}
\end{center}
\end{figure}

In this section, we investigate the sensitivity of the $91.2$ GeV CEPC to light, long-lived photophobic ALPs through displaced-vertex signals. Specifically, we consider the single-production processes of the photophobic ALP, $e^+ e^- \to a \gamma$, $a \nu \bar{\nu}$, $a e^+ e^-$, and $a jj$. The Feynman diagrams for these processes and their corresponding production cross sections at the CEPC are presented in Figs.~\ref{fig:3} and \ref{fig:4}, respectively. The numerical results shown in Fig.~\ref{fig:4} are obtained after applying a set of basic kinematic cuts. We require the transverse momenta of photons and leptons to satisfy $p_T^{\gamma,\ell} > 10~\mathrm{GeV}$, while jets are required to satisfy $p_T^{j} > 20~\mathrm{GeV}$. The absolute pseudorapidities of photons, leptons, and jets are required to satisfy $|\eta_{\gamma,\ell}| < 2.5$ and $|\eta_{j}| < 5$, respectively. In addition, the angular separations between the two leptons and between the two jets are required to satisfy $\Delta R_{\ell\ell} > 0.4$ and $\Delta R_{jj} > 0.4$, where $\Delta R$ is defined as $\Delta R =\sqrt{(\Delta \phi)^2 + (\Delta \eta)^2}$.

As shown in Fig.~\ref{fig:4}, the photon-associated production process $e^+ e^- \to a \gamma$ clearly dominates. For ALPs with masses in the range of $1$--$10$ GeV and $g_{aWW} = 1 \times 10^{-3}$ TeV$^{-1}$, its production cross section varies from $1.746 \times 10^{-5}$ to $1.683 \times 10^{-5}$ pb, which is approximately three orders of magnitude larger than those of the other processes. For this process, the mediator can be either a photon or a $Z$ boson. However, in the photophobic ALP scenario, the effective $a\gamma\gamma$ coupling is generated only at the loop level. Consequently, the photon-mediated contribution accounts for only about $0.1\%$ of the total production cross section and can be safely neglected. The $a\gamma$ final state is therefore predominantly produced via the on-shell decay $Z\to a\gamma$ at the $Z$ pole of the CEPC. We thus focus on the production process $e^+ e^- \to Z \to a\gamma$ and investigate the prospects for detecting ALP signals at the CEPC.

We now focus on the production process $e^+ e^- \to Z \to a \gamma $, in which the ALP subsequently decays into a pair of displaced charged leptons within the main tracking system of the CEPC at $\sqrt{s}=91.2 $ GeV and $\mathcal{L}=$ $100$ ab$^{-1}$. We consider ALPs with masses in the range of $1-10$ GeV, focusing on the displaced dimuon channel $a \to \mu^+\mu^-$ for $m_a=1-4$ GeV and the displaced ditau channel $a \to \tau^+\tau^-$ for $m_a=4-10$ GeV. Since tau leptons are unstable, they can decay either leptonically or hadronically, with the hadronic modes having larger branching fractions. We therefore focus on the 1-prong hadronic tau decay modes as representative examples, in which each $\tau$ decays via $\tau^{\pm}\to\pi^{\pm}\nu_\tau$ or $\tau^{\pm}\to\pi^{\pm}\pi^0\nu_\tau$. The corresponding signals are characterized by $\mu^+ \mu^- \gamma$ for the dimuon channel and $\tau_h^+ \tau_h^- \slashed{E}_T \gamma$ for the ditau channel.

For further calculations, we use the \verb"FeynRules" package~\cite{Alloul:2013bka} to generate the UFO model file for the effective Lagrangian. For the numerical results and event generation, we use Monte Carlo~(MC) simulations with \verb"MadGraph5_aMC@NLO"~\cite{Alwall:2014hca} to generate the signal and background events at the parton level with the following basic cuts:
\begin{equation}\label{eq:10}
\begin{split}
\hspace{-3em} \text{For the dimuon channel}: ~~& p_T^{\ell} > 5~\text{GeV}, \quad |\eta_{\ell}| < 2.5,~\text{with}~\ell = \mu, \\
    & p_T^\gamma > 10~\text{GeV}, \quad |\eta_\gamma| < 2.5, \\
    & \Delta R_{\ell\ell} > 0.2, \quad \Delta R_{\ell\gamma} > 0.2.\\
\hspace{-3em} \text{For the ditau channel}: ~~ & p_T^\gamma > 10~\text{GeV}, \quad |\eta_\gamma| < 2.5.
\end{split}
\end{equation}
Here, we use the \verb"PYTHIA8" program~\cite{Sjostrand:2014zea} to perform parton showering. Detector effects are simulated using the simplified fast simulation~(SFS) framework~\cite{Araz:2020lnp,Araz:2021akd} embedded in \texttt{MadAnalysis5}~\cite{Conte:2012fm,Conte:2014zja,Conte:2018vmg}. The detector response, including reconstruction efficiencies and momentum and energy smearing, is modeled using parameter settings based on the official CEPC detector card~\texttt{delphes\_card\_CEPC.tcl}~\footnote{\url{https://github.com/delphes/delphes/blob/master/cards/delphes_card_CEPC.tcl}} from \texttt{DELPHES}~\cite{deFavereau:2013fsa}. The SFS particle propagator module simulates charged-particle propagation in a homogeneous magnetic field, enabling the reconstruction of displacement-related observables. The corresponding analytical expressions are provided in Appendix~A of Ref.~\cite{Araz:2021akd}. In our analysis, the magnetic field strength is set to $3.5$~T, and the tracker radius is taken to be $1.8$~m. No additional detector smearing associated with finite spatial or tracking resolutions is applied to these displacement-related observables. Because the LLP signals considered in this work typically have decay lengths much larger than the expected detector spatial resolution, omitting such effects is expected to have only a minor impact on our phenomenological results. The subsequent kinematic and cut-based analysis is performed using C++ in the expert mode of \verb|MadAnalysis5|.

\subsection{The $\mu^+ \mu^- \gamma$ signal }\label{subsec1}

\begin{figure}[H]
\begin{center}
\centering\includegraphics [scale=0.33] {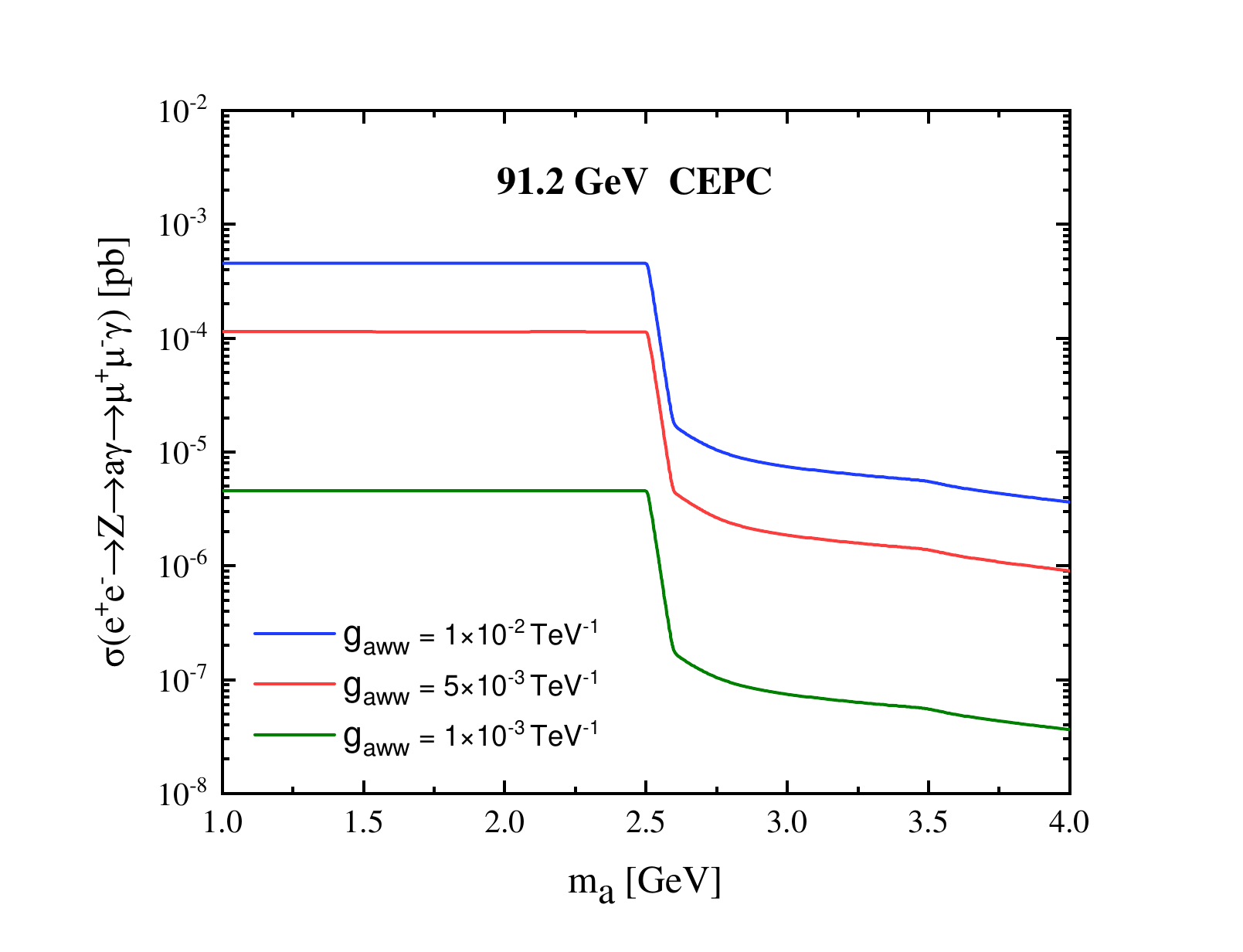}
\caption{Production cross sections for the $\mu^+ \mu^- \gamma$ signal as functions of the ALP mass $m_a$ for several representative values of the coupling $g_{aWW}$ at the CEPC operating at 91.2 GeV.}
\label{fig:5}
\end{center}
\end{figure}

\begin{figure}[H]
\begin{center}
\subfigure[]{\includegraphics [scale=0.55] {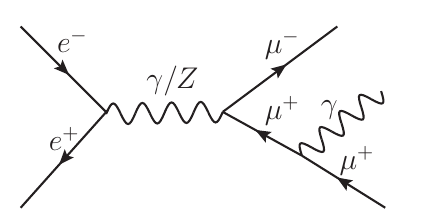}}
\subfigure[]{\includegraphics [scale=0.55] {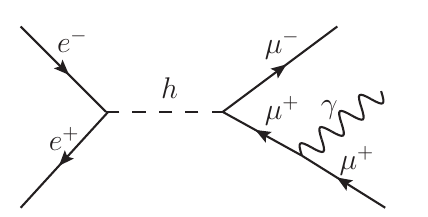}}
\subfigure[]{\includegraphics [scale=0.55] {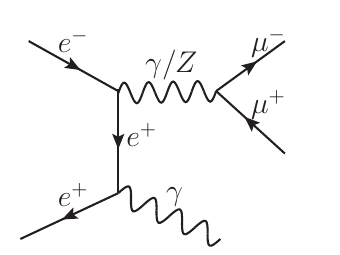}}
\subfigure[]{\includegraphics [scale=0.55] {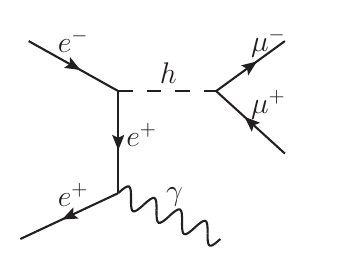}}
\subfigure[]{\includegraphics [scale=0.47] {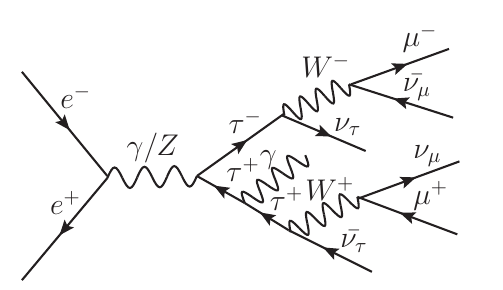}}
\subfigure[]{\includegraphics [scale=0.47] {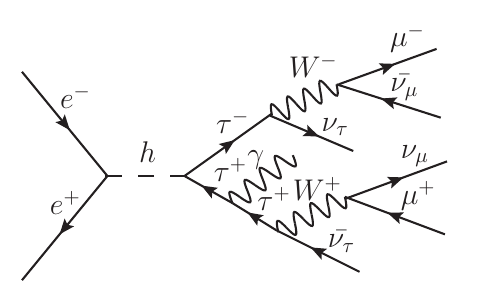}}
\subfigure[]{\includegraphics [scale=0.47] {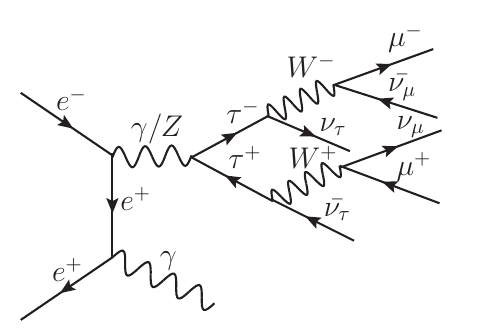}}
\subfigure[]{\includegraphics [scale=0.47] {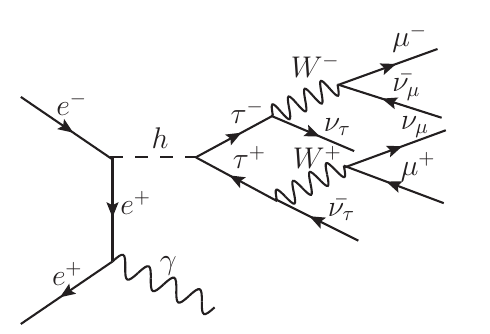}}
\caption{The Feynman diagrams correspond to the SM background processes $e^+e^- \to \mu^+\mu^-\gamma$~(a--d), $e^+e^- \to \tau^+\tau^-$, in which one tau lepton radiates a photon before decaying into $\mu \nu \bar{\nu}$~(e,f), and $e^+e^- \to \tau^+\tau^-\gamma$, in which both tau leptons subsequently decay into $\mu \nu \bar{\nu}$~(g,h).}
\label{fig:6}
\end{center}
\end{figure}

We now focus on the $\mu^+ \mu^- \gamma$ signal for ALPs with masses in the range of $1-4$ GeV.
The corresponding production cross sections at the CEPC are shown in Fig.~\ref{fig:5}. In general, the production cross section decreases as $m_a$ increases. A noticeable suppression occurs around $m_a \simeq 2.5~\mathrm{GeV}$ because of the kinematic opening of the $a \to c\bar{c}$ decay channel, which reduces the branching ratio of $a \to \mu^+\mu^-$ and consequently suppresses the $\mu^+\mu^-\gamma$ signal.
After applying the basic cuts described above, the production cross sections range from $4.549 \times 10^{-4}$ to $3.625 \times 10^{-6}$ pb, from $1.137 \times 10^{-4}$ to $9.053 \times 10^{-7}$ pb, and from $4.550 \times 10^{-6}$ to $3.621 \times 10^{-8}$ pb for $g_{aWW}=1 \times 10^{-2}$ TeV$^{-1}$, $5 \times 10^{-3}$ TeV$^{-1}$, and $1 \times 10^{-3}$ TeV$^{-1}$, respectively. The dominant SM background arises from the process $e^+e^- \to \mu^+\mu^-\gamma$, whereas subleading contributions arise from $e^+e^- \to \tau^+\tau^-\gamma$, with each tau subsequently decaying into $\mu \nu \bar{\nu}$, and from $e^+e^- \to \tau^+\tau^-$, with one of the taus radiating a photon before decaying leptonically. The corresponding Feynman diagrams for these background processes are shown in Fig.~\ref{fig:6}. The production cross section of the $\mu^+\mu^-\gamma$ final state is approximately $37.121$ pb, and the total production cross section of the SM background is approximately $38.389$ pb. In addition, semileptonic and dileptonic decays of heavy-flavor hadrons, such as $b$- and $c$-hadrons, may also contribute to the SM background~\cite{ALEPH:2005ab}. Owing to the relatively long lifetimes of heavy-flavor hadrons, their semileptonic and dileptonic decays can produce displaced muons, thereby constituting a potential background for displaced-vertex searches. To evaluate these heavy-flavor background contributions, we simulate the processes $e^+e^- \rightarrow \gamma^{*}/Z \rightarrow c\bar{c}, b\bar{b}$ at the $Z$ pole using \texttt{PYTHIA}~\cite{Sjostrand:2014zea}. Event generation is performed with the settings \texttt{WeakSingleBoson:ffbar2gmZ=on}, \texttt{23:onMode=off}, and \texttt{23:onIfAny=4 5}. Initial- and final-state radiation, heavy-quark fragmentation and hadronization, and subsequent decays of unstable heavy-flavor hadrons are included in the simulation. The total production cross section for the heavy-flavor background is approximately $1.180\times10^{4}$ pb, and the corresponding Feynman diagrams are not shown for brevity. In addition to these physics backgrounds, instrumental backgrounds, such as pile-up interactions, vertices originating from dense detector regions, and random-track crossings, may in principle be present. A quantitative assessment of these effects requires dedicated detector-level simulations and specialized reconstruction techniques, which are beyond the scope of this study. Accordingly, the results presented in this paper are obtained under the simplifying assumption that these instrumental backgrounds are negligible.

\begin{figure}[H]
\begin{center}
\subfigure[]{\includegraphics [scale=0.35] {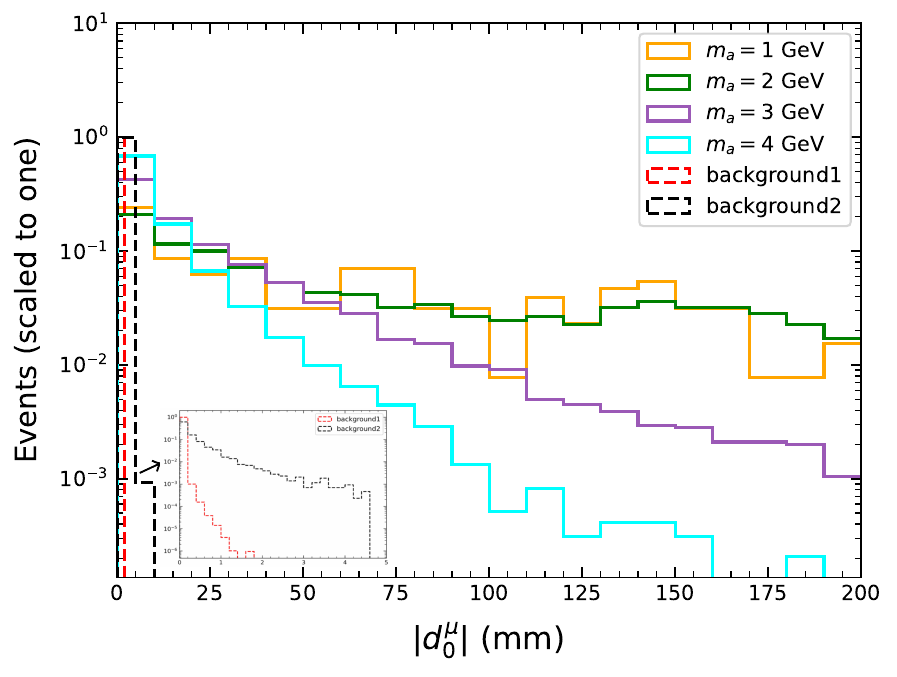}}
\hspace{0.1in}
\subfigure[]{\includegraphics [scale=0.35] {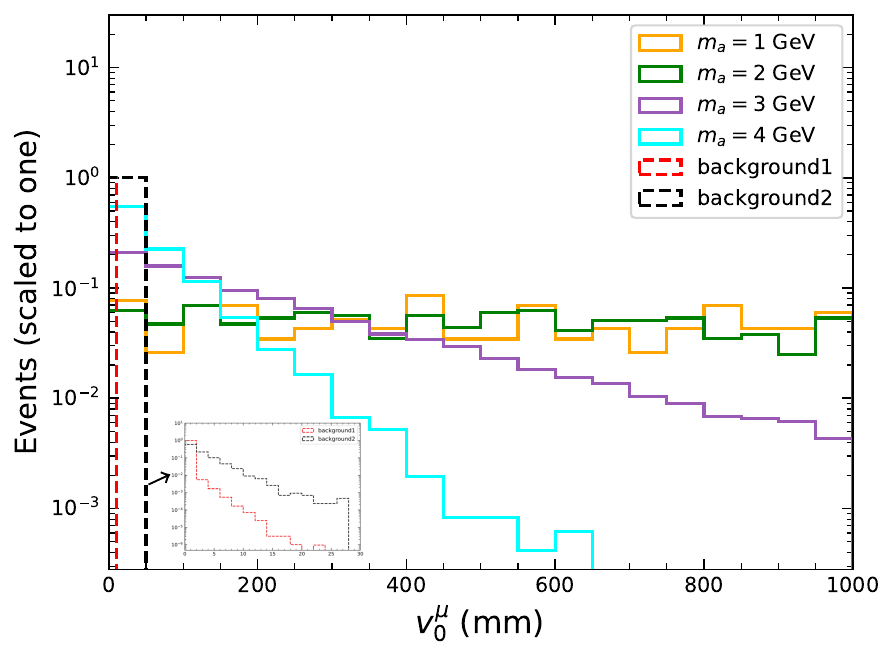}}
\hspace{0.1in}
\subfigure[]{\includegraphics [scale=0.35] {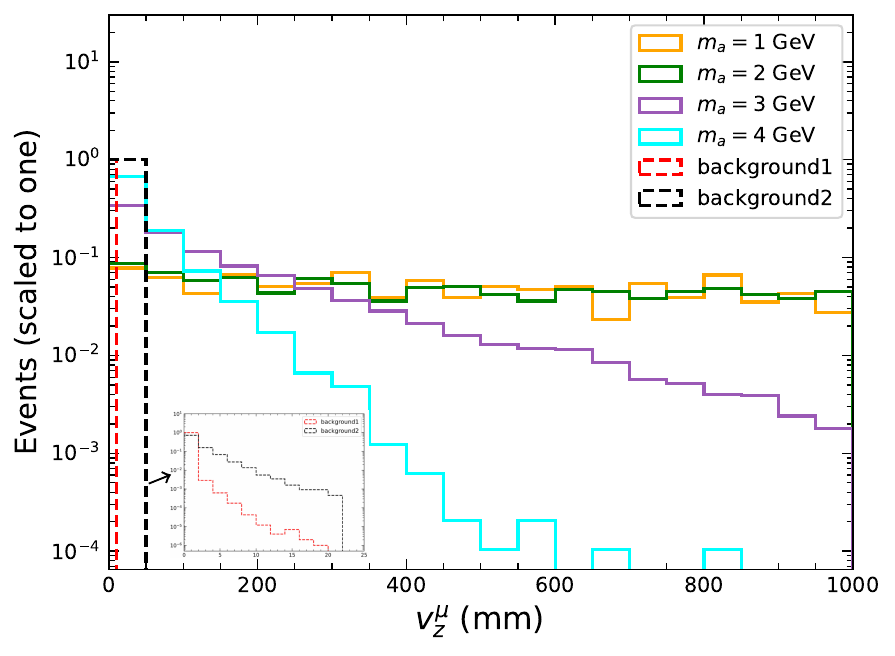}}
\subfigure[]{\includegraphics [scale=0.35] {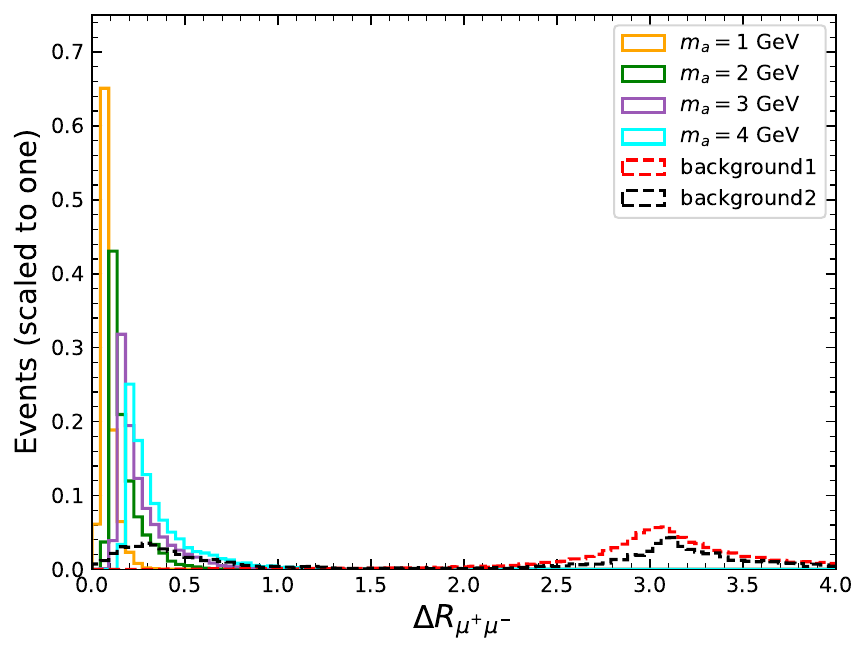}}
\subfigure[]{\includegraphics [scale=0.35] {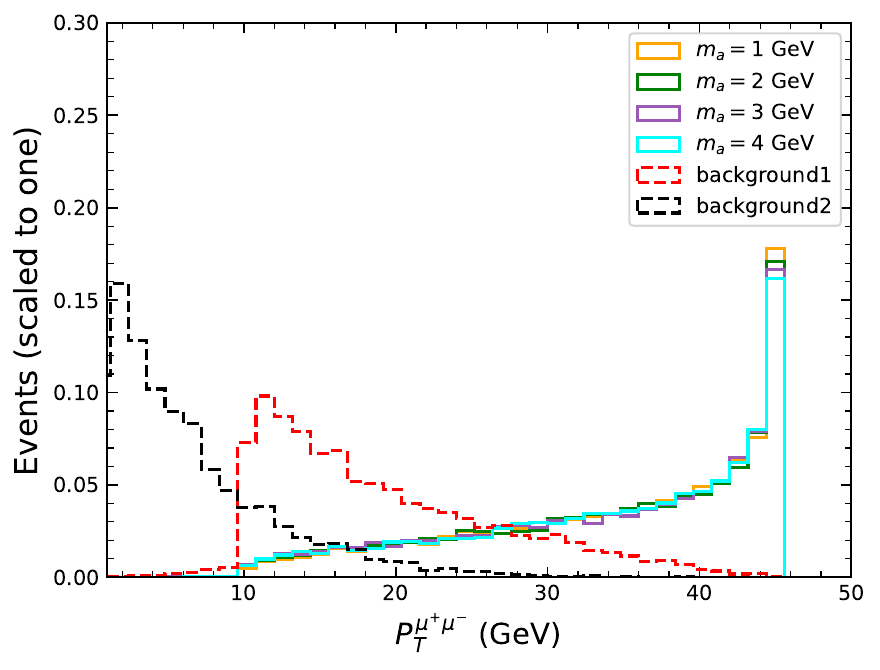}}
\caption{The normalized distributions of $|d_0^{\mu}|$, $v_0^{\mu}$, $v_z^{\mu}$, $\Delta R_{\mu^+\mu^-}$, and $P_T^{\mu^+\mu^-}$ for the $\mu^+ \mu^- \gamma$ signal at selected ALP mass benchmark points with $g_{aWW}=5 \times 10^{-3}$ TeV$^{-1}$, as well as for the SM background, at the CEPC operating at 91.2 GeV with $\mathcal{L}=$ $100$ ab$^{-1}$.}
\label{fig:7}
\end{center}
\end{figure}

Several vertex-displacement observables are used to distinguish the signal from the SM background, including $|d_0^{\mu}|$, $v_0^{\mu}$, and $v_z^{\mu}$. The observable $|d_0^{\mu}|$ denotes the transverse impact parameter of the charged muon, defined as the distance of closest approach between the helical trajectory of the muon track and the beam axis in the transverse plane. The observables $v_0^{\mu}$ and $v_z^{\mu}$ correspond to the transverse and longitudinal distances, respectively, of the muon production vertex relative to the interaction point. In addition to these displacement observables, we consider the angular separation $\Delta R_{\mu^+\mu^-}$ and the transverse momentum $P_T^{\mu^+\mu^-}$ of the reconstructed ALP to further discriminate the signal from the SM background. The normalized distributions of these observables for the $\mu^+ \mu^- \gamma$ signal with ALP masses $m_a = 1$, $2$, $3$, and $4$ GeV and fixed coupling $g_{aWW}= 5 \times 10^{-3}$ TeV$^{-1}$, as well as for the SM background at the $91.2$ GeV CEPC with $\mathcal{L} = 100$ ab$^{-1}$, are shown in Fig.~\ref{fig:7}. The colored solid curves correspond to the signal events, whereas the red and black dashed curves represent the prompt SM background and the heavy-flavor background originating from semileptonic and dileptonic decays of $b$- and $c$-hadrons, respectively.

As shown in Fig.~\ref{fig:7}(a-c), the signal events exhibit broad distributions with extended tails in $|d_0^{\mu}|$, $v_0^{\mu}$, and $v_z^{\mu}$, reflecting the displaced decay of a light, long-lived ALP. In contrast, the prompt SM background mainly arises from promptly decaying particles and is therefore concentrated at small values of these observables. The heavy-flavor background is dominated by semileptonic and dileptonic decays of $b$- and $c$-hadrons. Owing to their relatively long lifetimes, these heavy-flavor hadrons produce muons with larger values of $|d_{0}^{\mu}|$, $v_{0}^{\mu}$, and $v_{z}^{\mu}$ than those from the prompt SM background. However, these displacement observables remain substantially smaller than those of the signal, in which the displaced muons are produced in the decays of a long-lived ALP. As shown in Fig.~\ref{fig:7}(d), the two muons from the long-lived ALP decay tend to be collimated, whereas their angular separation in background events exhibits a much broader distribution. Figure~\ref{fig:7}(e) shows the $P_T^{\mu^+\mu^-}$ distribution. For the signal events, the ALP is produced via the two-body decay $Z \to \gamma a$ and is therefore typically highly boosted, resulting in a dimuon system with relatively large transverse momentum. By contrast, the dimuon pair in background events generally carries smaller transverse momentum, yielding a distribution concentrated in the low-$P_T$ region.

\begin{table}[H]
\begin{center}
\setlength{\tabcolsep}{3mm}{
\caption{Optimized cuts for the $\mu^+ \mu^- \gamma$ signal and SM background for $1~\mathrm{GeV} \leq m_a \leq 4~\mathrm{GeV}$ at the $91.2$ GeV CEPC with $\mathcal{L}=100$ ab$^{-1}$.}
\label{tab1}
\resizebox{0.75\textwidth}{!}{
\begin{tabular}
[c]{c|c c c c c}\hline \hline
\multicolumn{1}{c|}{\textbf{Optimized cuts}} & \textbf{Selection criteria} \\
\hline
\makecell[l]{Cut 1: the multiplicity, transverse momentum and pseudorapidity \\ of the muons in the final states } &
\makecell[c]{$N_{\mu^{+}} \geq 1$, $N_{\mu^{-}} \geq 1$, \\ $p_T^{\mu} > 5~\text{GeV}$, $|\eta_{\mu}| < 2.5$} \\
\makecell[l]{Cut 2: the transverse impact parameter} &
\makecell[c]{$|d_{\textnormal{0}}^{\mu}| > 2~\mathrm{mm}$} \\
\makecell[l]{Cut 3: the transverse and longitudinal distance} &
\makecell[c]{$0.1~\mathrm{m} < v_{\textnormal{0}}^{\mu} < 1.8~\mathrm{m}$, $v_z^{\mu} < 2.35~\mathrm{m}$} \\
\makecell[l]{Cut 4: the angular separation between the two charged muons} &
\makecell[c]{$\Delta R_{\mu^+\mu^-} < 1.0$} \\
\makecell[l]{Cut 5: the transverse momentum of reconstructed ALP} &
\makecell[c]{$p_T^{\mu^{+}\mu^{-}} > 25~\mathrm{GeV}$} \\
\hline \hline
\end{tabular}}}
\end{center}
\end{table}

Based on the characteristic features of these distributions, the optimized cuts for identifying the $\mu^+ \mu^- \gamma$ signal and suppressing the SM background are summarized in Table~\ref{tab1}. Requirements on the multiplicity, transverse momentum, and pseudorapidity of the final-state muons serve as the first-step filter. These cuts include fiducial requirements on the displacement observables: $|d_0^{\mu}| > 2~\text{mm}$, $0.1~\text{m} < v_0^{\mu} < 1.8~\text{m}$, and $v_z^{\mu} < 2.35~\text{m}$. These requirements are chosen according to the detector geometry and previous displaced-vertex studies~\cite{CEPCStudyGroup:2018ghi,Cheung:2019qdr,Cao:2023smj}, ensuring that the ALP decays occur within the sensitive region of the main tracking system. In Tables~\ref{tab2} and \ref{tab3}, we summarize the cumulative selection efficiency $\epsilon$ and production cross section $\sigma$ of the $\mu^+ \mu^- \gamma$ signal for representative ALP mass points with $g_{aWW}=5\times10^{-3}$ TeV$^{-1}$, as well as those of the SM background, after the step-by-step optimized cuts employed at the $91.2$ GeV CEPC with $\mathcal{L}=100$ ab$^{-1}$.

As shown in Tables~\ref{tab2} and \ref{tab3}, the cumulative selection efficiency of the signal ranges from $8.00\times10^{-3}$ to $1.63\times10^{-1}$ for ALP masses between 1 and 4~GeV, while no SM background events are observed after applying the full set of optimized cuts to the $10^8$ simulated Monte Carlo events. Based on the available Monte Carlo statistics, we conservatively estimate an upper limit on the background efficiency of $\epsilon < 10^{-8}$. Consequently, the SM background is expected to be strongly suppressed within the present simulation framework. Although a small number of background events may appear in the extreme tails of the distributions if substantially larger Monte Carlo samples are generated, a reliable evaluation of such rare fluctuations would require prohibitively large computational resources and is therefore left for future work.

\begin{table}[H]\tiny
	\centering{
\caption{The cumulative selection efficiency $\epsilon$ and production cross section $\sigma$ of the $ \mu^+ \mu^- \gamma$ signal are shown for representative ALP mass points with $g_{aWW}=5\times10^{-3}$ TeV$^{-1}$ after the sequential optimized cuts applied at the $91.2$ GeV CEPC with $\mathcal{L}=$ $100$ ab$^{-1}$.$~$\label{tab2}}
		\newcolumntype{C}[1]{>{\centering\let\newline\\\arraybackslash\hspace{50pt}}m{#1}}
		\begin{tabular}{m{2cm}<{\centering}|m{2.3cm}<{\centering} m{2.3cm}<{\centering} m{2.3cm}<{\centering}  m{2.3cm}<{\centering} m{2.3cm}<{\centering}}\hline \hline
      \multirow{2}{*}{Cuts} & \multicolumn{4}{c}{Cross sections~[pb] (selection efficiencies) for the signal}\\
     \cline{2-5}
     & $m_a=1$ GeV  & $m_a=2$ GeV  & $m_a=3$ GeV  & $m_a=4$ GeV  \\ \hline
     Generator:  & \makecell{$1.137\times10^{-4}$} & \makecell{$1.138\times10^{-4}$} & \makecell{$1.813\times10^{-6}$} & \makecell{$9.053\times10^{-7}$} \\ \hline
     Cut 1:  & \makecell{$1.064\times10^{-4}$\\($9.36\times10^{-1}$)} & \makecell{$1.050\times10^{-4}$\\($9.22\times10^{-1}$)} & \makecell{$1.667\times10^{-6}$\\($9.20\times10^{-1}$)} & \makecell{$8.318\times10^{-7}$\\($9.19\times10^{-1}$)} \\ \hline
     Cut 2:  & \makecell{$1.063\times10^{-4}$\\($9.34\times10^{-1}$)} & \makecell{$1.042\times10^{-4}$\\($9.15\times10^{-1}$)} & \makecell{$1.096\times10^{-6}$\\($6.05\times10^{-1}$)} & \makecell{$3.457\times10^{-7}$\\($3.82\times10^{-1}$)} \\  \hline
     Cut 3:  & \makecell{$1.135\times10^{-6}$\\($1.00\times10^{-2}$)} & \makecell{$5.462\times10^{-6}$\\($4.80\times10^{-2}$)} & \makecell{$9.027\times10^{-7}$\\($4.98\times10^{-1}$)} & \makecell{$1.574\times10^{-7}$\\($1.74\times10^{-1}$)} \\  \hline
     Cut 4:  & \makecell{$1.135\times10^{-6}$\\($1.00\times10^{-2}$)} & \makecell{$5.462\times10^{-6}$\\($4.80\times10^{-2}$)} & \makecell{$9.027\times10^{-7}$\\($4.98\times10^{-1}$)} & \makecell{$1.574\times10^{-7}$\\($1.74\times10^{-1}$)} \\  \hline
     Cut 5:  & \makecell{$9.091\times10^{-7}$\\($8.00\times10^{-3}$)} & \makecell{$3.411\times10^{-6}$\\($3.00\times10^{-2}$)} & \makecell{$7.611\times10^{-7}$\\($4.20\times10^{-1}$)} & \makecell{$1.474\times10^{-7}$\\($1.63\times10^{-1}$)} \\
     \hline \hline
	\end{tabular}}	
\end{table}

The number of signal events passing the optimized selection criteria is given by $N_{sig} = \epsilon \times \sigma \times \mathcal{L}$. After applying these criteria, the SM background is negligible in our analysis. We therefore adopt 3 signal events as the benchmark for the $95\%$ confidence level~(C.L.) exclusion limit. To derive the projected $95\%$ C.L. sensitivity, we scan the parameter space $(m_a, g_{aWW})$ and identify the region satisfying $N_{sig} \geq 3$. The resulting $95\%$ C.L. sensitivity of the 91.2 GeV CEPC with $\mathcal{L}=$ $100$ ab$^{-1}$ to light, long-lived photophobic ALPs via the $ \mu^+ \mu^- \gamma$ signal is shown by the purple area in Fig.~\ref{fig:11}.

\begin{table}[H]\tiny
	\centering{
\caption{Same as Table~\ref{tab2}, but for the SM background.\label{tab3}}
		\newcolumntype{C}[1]{>{\centering\let\newline\\\arraybackslash\hspace{50pt}}m{#1}}
		\begin{tabular}{m{2cm}<{\centering}|m{3.2cm}<{\centering} m{3.2cm}<{\centering}}\hline \hline
      \multirow{2}{*}{Cuts} & \multicolumn{2}{c}{Cross sections~[pb] (selection efficiencies) for the SM background}\\
     \cline{2-3}
     & background1  & background2  \\ \hline
     Generator:  & \makecell{$3.839\times10^{1}$} & \makecell{$1.180\times10^{4}$} \\ \hline
     Cut 1:  & \makecell{$3.808\times10^{1}$\\($9.92\times10^{-1}$)} & \makecell{$2.536\times10^{2}$\\($2.15\times10^{-2}$)}\\ \hline
     Cut 2:  & \makecell{$<3.839\times10^{-7}$\\($< 1.00\times10^{-8}$)} & \makecell{$2.986\times10^{-1}$\\($2.53\times10^{-5}$)} \\  \hline
     Cut 3:  & \makecell{$<3.839\times10^{-7}$\\($< 1.00\times10^{-8}$)} & \makecell{$<1.180\times10^{-4}$\\($< 1.00\times10^{-8}$)} \\  \hline
     Cut 4:  & \makecell{$<3.839\times10^{-7}$\\($< 1.00\times10^{-8}$)} & \makecell{$<1.180\times10^{-4}$\\($< 1.00\times10^{-8}$)} \\  \hline
     Cut 5:  & \makecell{$<3.839\times10^{-7}$\\($< 1.00\times10^{-8}$)} & \makecell{$<1.180\times10^{-4}$\\($< 1.00\times10^{-8}$)} \\
     \hline \hline
	\end{tabular}}	
\end{table}

\subsection{The $\tau_h^+ \tau_h^- \slashed{E}_T \gamma$ signal}\label{subsec2}

\begin{figure}[H]
\begin{center}
\centering\includegraphics [scale=0.35] {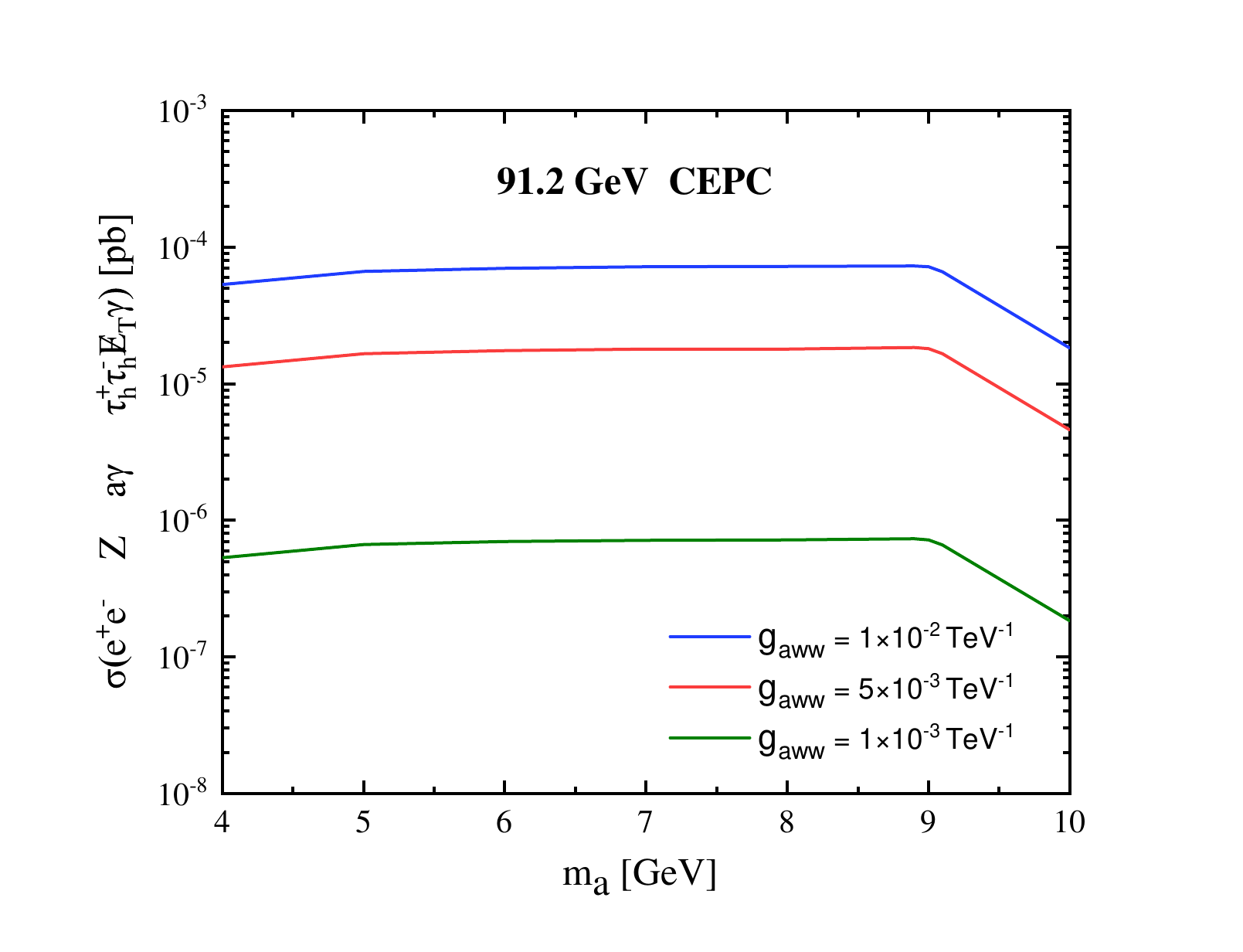}
\caption{Same as Fig.~\ref{fig:5}, but for the $\tau_h^+ \tau_h^- \slashed{E}_T \gamma$ signal.}
\label{fig:8}
\end{center}
\end{figure}

\begin{figure}[H]
\begin{center}
\subfigure[]{\includegraphics [scale=0.6] {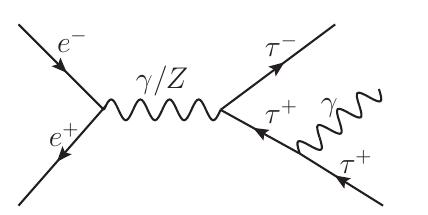}}
\subfigure[]{\includegraphics [scale=0.6] {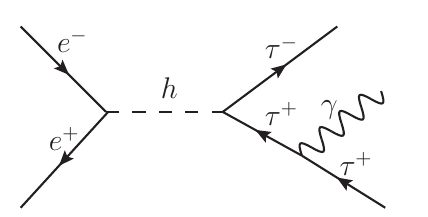}}
\subfigure[]{\includegraphics [scale=0.6] {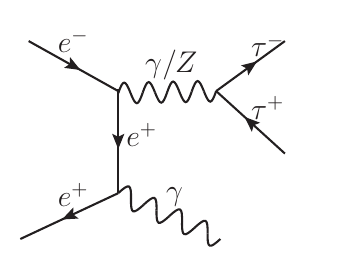}}
\subfigure[]{\includegraphics [scale=0.6] {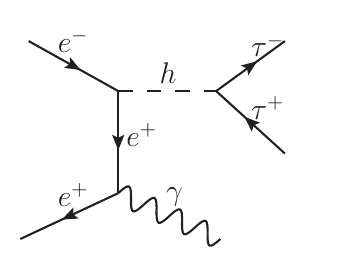}}
\caption{Feynman diagrams for the SM background process $e^+ e^- \to \tau^+ \tau^- \gamma$.}
\label{fig:9}
\end{center}
\end{figure}

In this subsection, we investigate the signal from the process $e^+ e^- \to Z \to a\gamma \to \tau^+ \tau^- \gamma$, focusing on subsequent hadronic tau decays. Each hadronic tau decay is reconstructed as a narrow tau jet containing a single charged track. Thus, the signal is characterized by two oppositely charged tau jets, missing transverse energy, and a photon, denoted as $\tau_h^+ \tau_h^- \slashed{E}_T \gamma$. The corresponding production cross sections at the CEPC as functions of the ALP mass $m_a$ for different values of $g_{aWW}$ are shown in Fig.~\ref{fig:8}. For $m_a$ in the range of $4-10$ GeV, the production cross sections range from $5.320 \times 10^{-5}$ to $1.835 \times 10^{-5}$ pb, from $1.330 \times 10^{-5}$ to $4.589 \times 10^{-6}$ pb, and from $5.320 \times 10^{-7}$ to $1.835 \times 10^{-7}$ pb for $g_{aWW} = 1 \times 10^{-2}$, $5 \times 10^{-3}$, and $1 \times 10^{-3}$ TeV$^{-1}$, respectively. These results are obtained after applying the basic cuts described above. In our numerical analysis, the dominant SM background arises from the processes $e^+ e^- \to \tau^+ \tau^- \gamma$, with both taus decaying via the 1-prong hadronic mode~($\tau^{\pm}\to\pi^{\pm}\nu_\tau$ or $\tau^{\pm}\to\pi^{\pm}\pi^0\nu_\tau$), and $e^+ e^- \to \tau^+ \tau^-$, with one tau radiating a photon before decaying hadronically. The corresponding Feynman diagrams are shown in Fig.~\ref{fig:9}, where the 1-prong hadronic tau decays are omitted for clarity. The production cross section is approximately $4.990$ pb.

\begin{figure}[H]
\begin{center}
\subfigure[]{\includegraphics [scale=0.35] {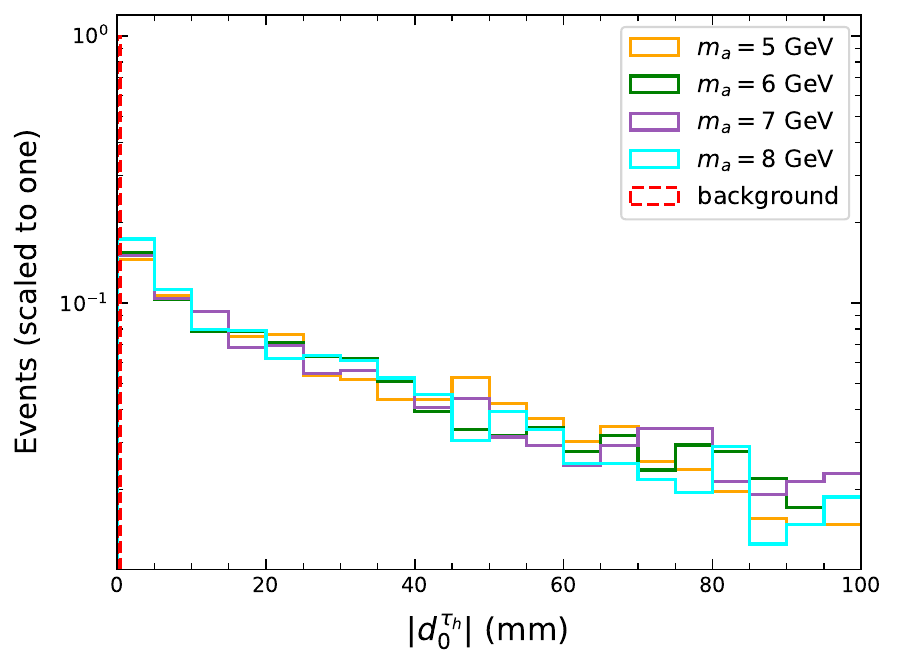}}
\hspace{0.1in}
\subfigure[]{\includegraphics [scale=0.35] {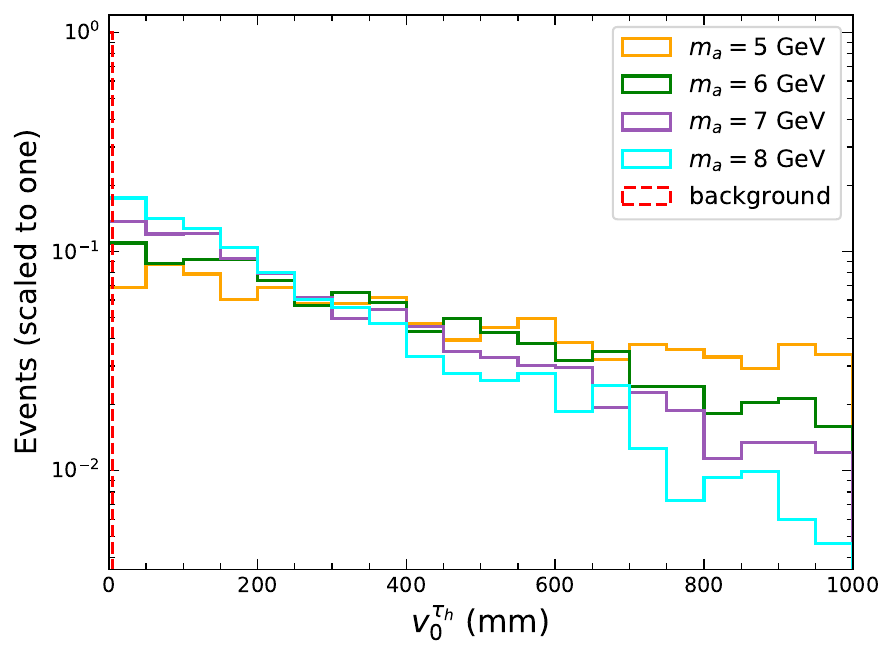}}
\hspace{0.1in}
\subfigure[]{\includegraphics [scale=0.35] {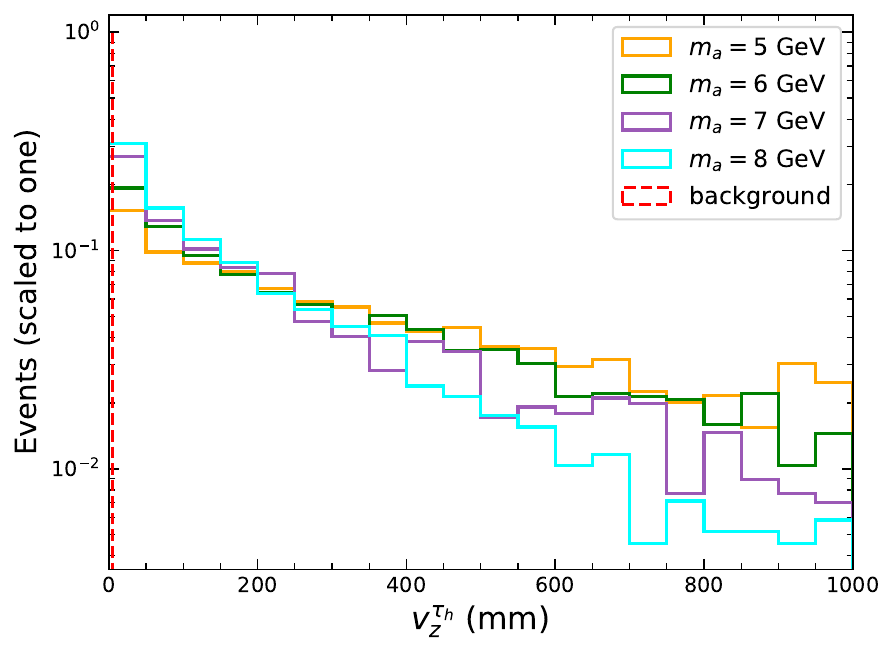}}
\subfigure[]{\includegraphics [scale=0.35] {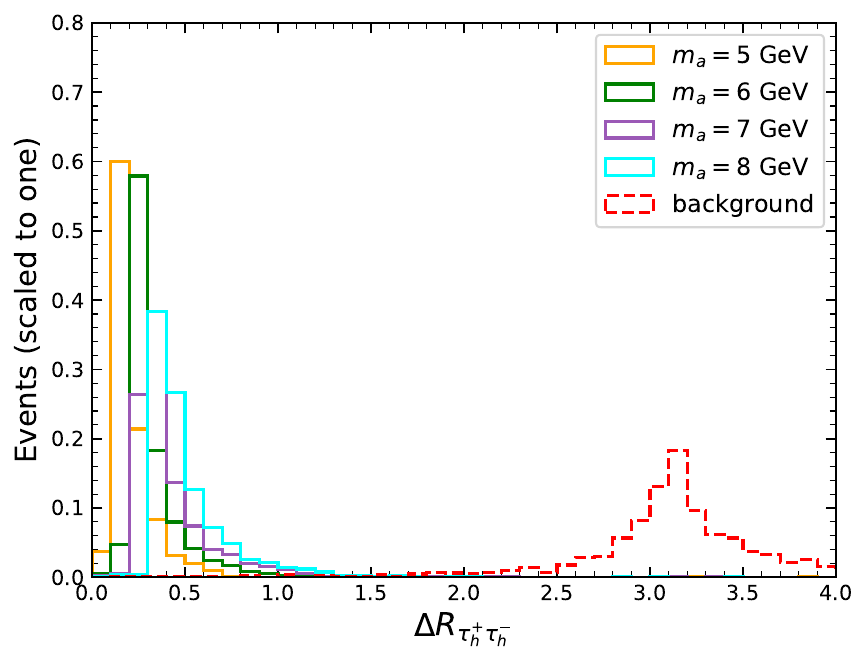}}
\subfigure[]{\includegraphics [scale=0.35] {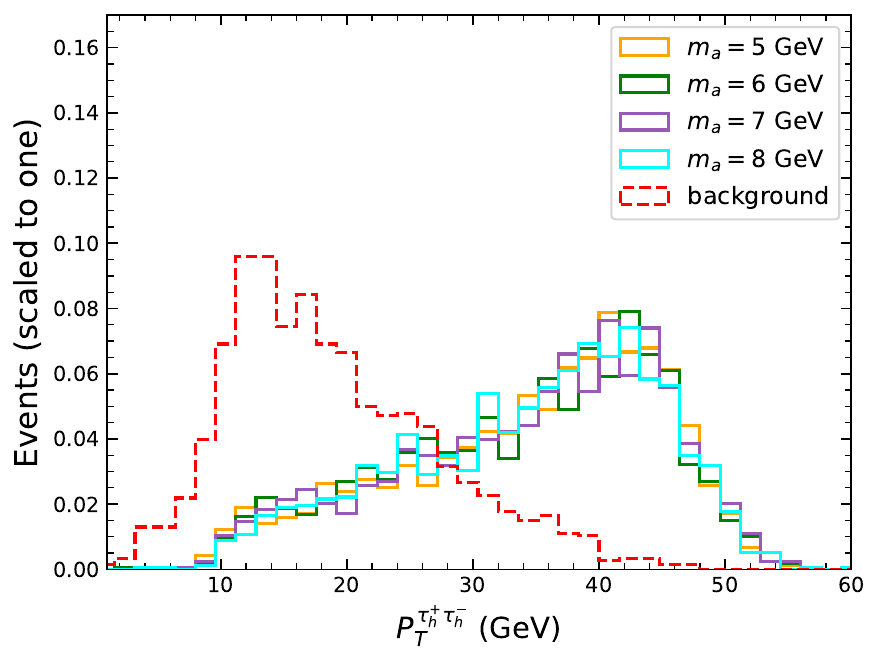}}
\caption{Same as Fig.~\ref{fig:7}, but for the observables $|d_0^{\tau_h}|$, $v_0^{\tau_h}$, $v_z^{\tau_h}$, $\Delta R _{\tau_h^+\tau_h^-}$, and $P_T^{\tau_h^+\tau_h^-}$ with $g_{aWW}=1 \times 10^{-3}$ TeV$^{-1}$.}
\label{fig:10}
\end{center}
\end{figure}

To effectively separate the signal from the SM background, we consider several observables similar to those used in the $\mu^+ \mu^- \gamma$ signal analysis, including $|d_0^{\tau_h}|$, $v_0^{\tau_h}$, $v_z^{\tau_h}$, $\Delta R_{\tau_h^+\tau_h^-}$, and $P_T^{\tau_h^+\tau_h^-}$. The normalized distributions for the $\tau_h^+ \tau_h^- \slashed{E}_T \gamma$ signal and the corresponding SM background at the 91.2 GeV CEPC with $\mathcal{L}=100~\rm ab^{-1}$ are shown in Fig.~\ref{fig:10}. The zoomed-in views of the SM background are similar to those of the prompt SM background~(red dashed curves)~shown in Fig.~\ref{fig:7}~(a-c) and are therefore omitted for brevity. As illustrated in this figure, the signal and background events can be clearly distinguished using these observables. Based on the features of these distributions, we apply optimized cuts similar to those used in the $\mu^+\mu^-\gamma$ signal analysis, but require $P_T^{\tau_h^+\tau_h^-} > 30$ GeV. After applying these optimized cuts, the background efficiency is estimated to be less than $10^{-8}$ and can therefore be considered negligible in the present analysis. Accordingly, the detailed cut-flow table for the SM background is omitted for brevity. Table~\ref{tab4} summarizes the cumulative selection efficiency $\epsilon$ and production cross section $\sigma$ for the $\tau_h^+ \tau_h^- \slashed{E}_T \gamma$ signal at the $91.2$ GeV CEPC with $\mathcal{L}=$ $100$ ab$^{-1}$, with $\epsilon$ ranging from $8.30\times10^{-2}$ to $7.60\times10^{-2}$ for $m_a=5-8$ GeV.

Assuming negligible SM background, we take $N_{sig}=3$ to estimate the expected 95$\%$ C.L. sensitivity of the $91.2$ GeV CEPC with $\mathcal{L}=100$ ab$^{-1}$ to light long-lived ALPs in the $\tau_h^+ \tau_h^- \slashed{E}_T \gamma$ channel, as shown by the pink region in Fig.~\ref{fig:11}. In this figure, we display only the parameter space for $m_a = 4-9$ GeV. For ALP masses above the $b\bar{b}$ threshold, the total decay width increases substantially, resulting in a much shorter lifetime. Consequently, the ALP decays are insufficiently displaced to satisfy the displaced-decay requirements within the main tracking system, leaving no accessible parameter space for long-lived ALPs in this mass region.

\begin{table}[H]\tiny
\begin{center}
\caption{As in Table~\ref{tab2}, but for the $\tau_h^+ \tau_h^- \slashed{E}_T \gamma$ signal with $g_{aWW}=1\times10^{-3}$ TeV$^{-1}$.$~$\label{tab4}}
\newcolumntype{C}[1]{>{\centering\let\newline\\\arraybackslash\hspace{50pt}}m{#1}}
\begin{tabular}{m{2.0cm}<{\centering}|m{2.3cm}<{\centering} m{2.3cm}<{\centering} m{2.3cm}<{\centering}  m{2.3cm}<{\centering} m{2.3cm}<{\centering}}\hline \hline
\multirow{2}{*}{Cuts} & \multicolumn{4}{c}{Cross sections~[pb] (efficiencies) for the signal} \\
\cline{2-5}
& $m_a=5$ GeV  & $m_a=6$ GeV  & $m_a=7$ GeV  & $m_a=8$ GeV  \\ \hline
Generator:  & \makecell{$6.653\times10^{-7}$} & \makecell{$6.998\times10^{-7}$} & \makecell{$7.145\times10^{-7}$} & \makecell{$7.187\times10^{-7}$} \\ \hline
Cut 1:  & \makecell{$1.084\times10^{-7}$\\($1.63\times10^{-1}$)} & \makecell{$1.112\times10^{-7}$\\($1.59\times10^{-1}$)} & \makecell{$1.129\times10^{-7}$\\($1.58\times10^{-1}$)} & \makecell{$1.130\times10^{-7}$\\($1.57\times10^{-1}$)} \\ \hline
Cut 2:       & \makecell{$9.778\times10^{-8}$\\($1.47\times10^{-1}$)} & \makecell{$9.721\times10^{-8}$\\($1.39\times10^{-1}$)} & \makecell{$9.646\times10^{-8}$\\($1.35\times10^{-1}$)} & \makecell{$9.271\times10^{-8}$\\($1.29\times10^{-1}$)} \\ \hline
Cut 3:       & \makecell{$8.316\times10^{-8}$\\($1.13\times10^{-1}$)} & \makecell{$8.258\times10^{-8}$\\($1.18\times10^{-1}$)} & \makecell{$7.931\times10^{-8}$\\($1.11\times10^{-1}$)} & \makecell{$7.331\times10^{-8}$\\($1.02\times10^{-1}$)} \\ \hline
Cut 4:       & \makecell{$8.316\times10^{-8}$\\($1.13\times10^{-1}$)} & \makecell{$8.258\times10^{-8}$\\($1.18\times10^{-1}$)} & \makecell{$7.931\times10^{-8}$\\($1.11\times10^{-1}$)} & \makecell{$7.115\times10^{-8}$\\($9.90\times10^{-2}$)} \\ \hline
Cut 5:       & \makecell{$5.522\times10^{-8}$\\($8.30\times10^{-2}$)} & \makecell{$5.738\times10^{-8}$\\($8.20\times10^{-2}$)} & \makecell{$5.716\times10^{-8}$\\($8.00\times10^{-2}$)} & \makecell{$5.462\times10^{-8}$\\($7.60\times10^{-2}$)} \\
\hline \hline
\end{tabular}
\end{center}
\end{table}

In Fig.~\ref{fig:11}, we present the projected $95\%$ C.L. sensitivities of the 91.2 GeV CEPC with $\mathcal{L}=$ $100$ ab$^{-1}$ to light long-lived photophobic ALPs. The purple and pink shaded regions denote the projected sensitivities obtained from the $\mu^+ \mu^- \gamma$ and $\tau_h^+ \tau_h^- \slashed{E}_T \gamma$ signals, respectively. For the $\mu^+ \mu^- \gamma$ signal, the CEPC is expected to probe couplings in the range $g_{aWW} \in [1.27 \times 10^{-3}, 6.80 \times 10^{-1}]~\mathrm{TeV}^{-1}$ for $m_a \in [1,4]~\mathrm{GeV}$. For the $\tau_h^+ \tau_h^- \slashed{E}_T \gamma$ signal, the prospective CEPC sensitivities can reach $g_{aWW} \in [7.00 \times 10^{-4}, 9.40 \times 10^{-3}]$ TeV$^{-1}$ for $m_a \in [4,9]$ GeV. Other excluded regions for photophobic ALPs from previous studies for $1$ GeV $\leq m_a \leq 100$ GeV are also shown as colored shaded regions in Fig.~\ref{fig:11}. These limits are adopted from the compilation of existing results in Refs.~\cite{Craig:2018kne,Aiko:2024xiv}. The gray and dark-gray regions represent the constraints on long-lived ALPs obtained by the LHCb collaboration through the processes $B^{0} \to K^{0*} a$ and $B^{\pm} \to K^{\pm} a$ with $a \to \mu^+ \mu^-$~\cite{LHCb:2016awg,LHCb:2015nkv}. The cyan region marked ``Charm" corresponds to the bounds on long-lived ALPs obtained from the CHARM experiment~\cite{CHARM:1985anb}. The yellow region marked ``LEP" denotes the constraints derived from the decays $Z \to \gamma a \to \gamma \gamma \slashed{E}_T$ and $Z \to \gamma a \to \gamma jj$ at the $Z$ pole at LEP~\cite{OPAL:1993ezs,L3:1992kcg}. The green region marked ``LHC" indicates the exclusion limits from various searches at the LHC~\cite{Craig:2018kne}, while the red region shows the bounds obtained from a dedicated Run-II 13 TeV LHC analysis targeting the process $pp \to a jj \to Z(\to \nu \bar{\nu}) \gamma jj$~\cite{Aiko:2024xiv}. Finally, the orange and blue regions indicate the projected sensitivities of the HL-LHC to light long-lived ALPs via the signals $ \pi^+ \pi^- \gamma \slashed{E}_T $ and $\ell^+ \ell^- \gamma \slashed{E}_T $, respectively~\cite{Yue:2025wkn}.

\begin{figure}[H]
\begin{center}
\centering\includegraphics [scale=0.45] {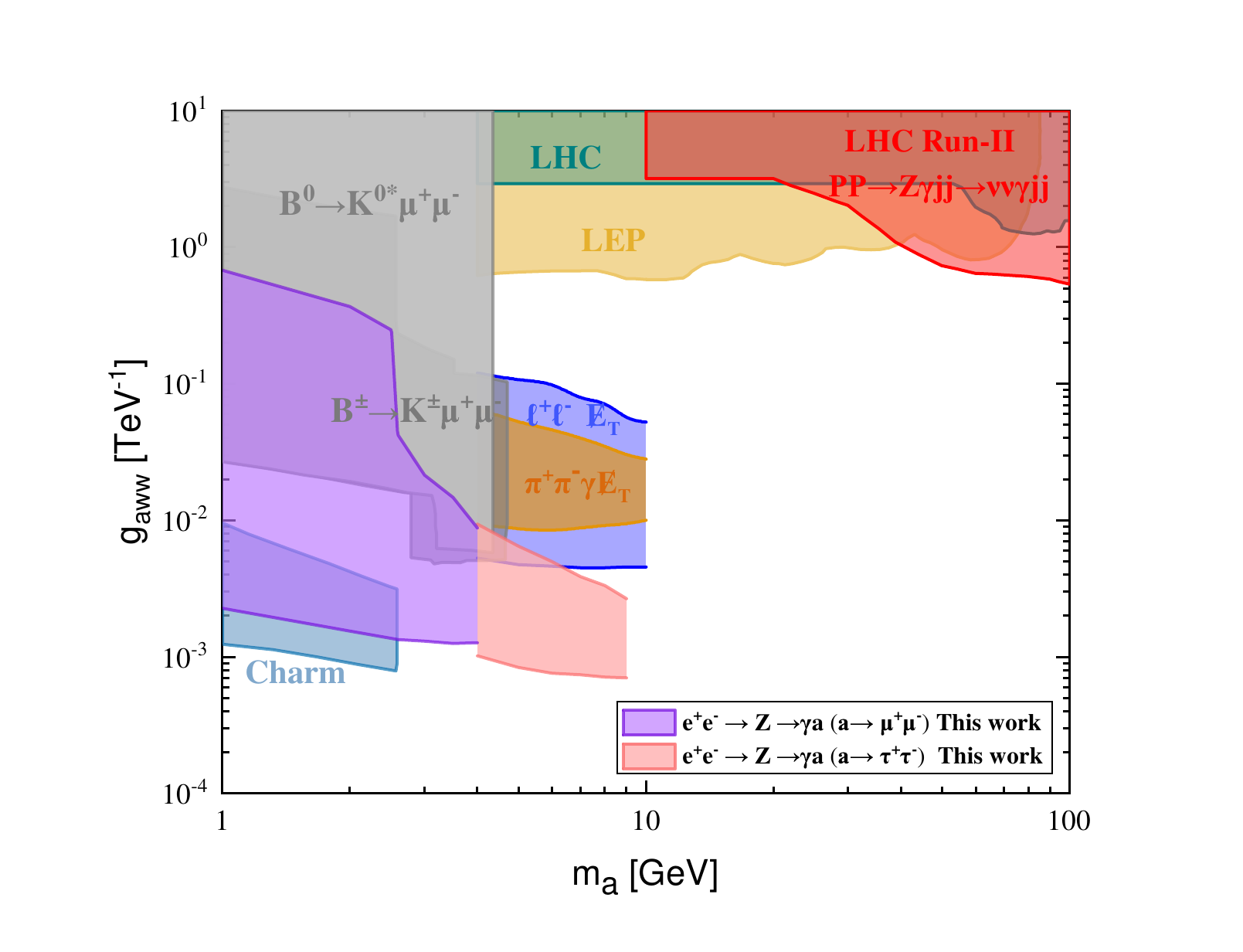}
\caption{Projected $95\%$ C.L. sensitivities at the CEPC for light, long-lived photophobic ALPs in the $\mu^+ \mu^- \gamma$~(purple region) and $\tau_h^+ \tau_h^- \slashed{E}_T \gamma$~(pink region) channels, compared with other excluded regions.}
\label{fig:11}
\end{center}
\end{figure}

Comparing our results with the exclusion regions from other experiments, we find that the 91.2 GeV CEPC with $\mathcal{L}=$ $100$ ab$^{-1}$ can probe long-lived photophobic ALPs with lower masses and weaker couplings via the $\mu^+ \mu^- \gamma$ and $\tau_h^+ \tau_h^- \slashed{E}_T \gamma$ signals, thereby accessing parameter regions beyond the reach of prompt-decay searches at LEP and the LHC. In addition, compared with the constraints on long-lived ALPs from rare meson decays obtained by the LHCb Collaboration and the CHARM experiment, the CEPC can provide complementary coverage of the light long-lived ALP parameter space. Furthermore, the projected sensitivity obtained in this work is complementary to the expected reach of the HL-LHC. In summary, these results indicate that the 91.2 GeV CEPC with $\mathcal{L}=$ $100$ ab$^{-1}$ has strong potential to explore light long-lived photophobic ALPs through displaced-vertex signals within the main tracking system.

\section{Conclusions and Discussions}

To date, a wide variety of new physics models have been proposed to address open problems in the SM and potential experimental anomalies, many of which predict new particles. Searching for possible signals of these particles is one of the primary goals of current and future collider experiments. As a well-motivated class of new particles, ALPs can give rise to rich phenomenology in both low- and high-energy experiments and therefore warrant extensive study. Many previous studies have focused on probing ALPs through their prompt decays or associated missing-energy signals.

In this paper, we have investigated the prospects for detecting light long-lived photophobic ALPs through displaced-vertex signals within the main tracking system of the CEPC with $\sqrt{s}=91.2$ GeV and $\mathcal{L} = 100$ ab$^{-1}$. We consider the production of a photophobic ALP in association with a photon, a lepton pair, or a jet pair. Among these production processes, we find that the photon-associated production mode has the largest production cross section. We therefore focus on the process $e^+ e^- \to Z \to \gamma a$, with the ALP $a$ subsequently decaying into a displaced pair of muons or tau leptons. We perform Monte Carlo simulations for the $\mu^+ \mu^- \gamma$ and $\tau_h^+ \tau_h^- \slashed{E}_T \gamma$ signals and derive the prospective $95\%$ C.L. sensitivities of the CEPC to light long-lived photophobic ALPs. For the $\mu^+ \mu^- \gamma$ signal, the CEPC can probe the ALP-$W$ boson coupling in the range $g_{aWW} \in [1.27 \times 10^{-3}, 6.80 \times 10^{-1}]~\mathrm{TeV}^{-1}$ for $m_a \in [1,4]$ GeV. For the $\tau_h^+ \tau_h^- \slashed{E}_T \gamma$ signal, the prospective sensitivities reach $g_{aWW} \in [7.00 \times 10^{-4}, 9.40 \times 10^{-3}]$ TeV$^{-1}$ for $m_a \in [4,9]$ GeV.
Comparing our results with exclusion regions from other experiments, we find that the CEPC can probe regions of the photophobic ALP parameter space characterized by lower masses and weaker couplings that remain challenging for prompt-decay searches at LEP and the LHC. In addition, compared with constraints from rare meson decays obtained by the LHCb collaboration and the CHARM experiment, the CEPC can provide complementary coverage of the light long-lived ALP parameter space. Furthermore, the projected sensitivity obtained in this work is complementary to the expected reach of the HL-LHC. Thus, we conclude that the CEPC has great potential to detect light long-lived photophobic ALPs via displaced-vertex signals.

In this paper, we primarily explore the potential of the CEPC for detecting light long-lived ALPs, taking advantage of its clean experimental environment, high integrated luminosity, and precise measurement capabilities. Other future lepton colliders, such as the FCC-ee, are also expected to offer promising opportunities for probing light long-lived ALPs. However, their center-of-mass energies, integrated luminosities, and detailed detector designs differ from those of the CEPC and therefore require dedicated analyses. In this study, we take the CEPC as a representative example and expect that our results may provide valuable insights into detecting light long-lived ALPs at the FCC-ee.

\section*{ACKNOWLEDGMENT}

This work was partially supported by the National Natural Science Foundation of China under Grant No. 12575106 and by the Cultivation Fund of Liaoning Normal University for Excellent Doctoral Dissertations~(No. YJSYB202501).

\end{document}